\documentclass[manuscript]{aastex}

\usepackage{txfonts}
\usepackage{longtable}
\usepackage{rotating}
\usepackage{natbib}
\usepackage{graphicx}
\usepackage{graphics}
\usepackage{psfrag}
\usepackage{amssymb}
\bibpunct{(}{)}{;}{a}{}{,}
\def\Teff{$T_{\mathrm{eff}}$}
\def\logg{\ensuremath{\log g}}

\def\kms{$\mathrm{km\,s}^{-1}$}

\def\loggf{log $gf$}

\def\llm{LLmodels}

\def\R{\ensuremath{R/R_{\odot}}}
\def\Ro{\ensuremath{R_{\odot}}}

\def\Mo{\ensuremath{M_{\odot}}}
\def\ebv{\ensuremath{E(B-V)}}

\def\ergscm{erg\,s$^{-1}$\,cm$^{-2}$}
\def\logR{\ensuremath{\log R^{\prime}_{\mathrm{HK}}}}

\shorttitle{Far-UV spectroscopy of the planet-hosting star WASP-13}
\shortauthors{Fossati et al.}

\begin{document}


\title{Far-UV spectroscopy of the planet-hosting star WASP-13: high-energy irradiance, distance, age, planetary mass-loss rate, and circumstellar environment}
\altaffiltext{1}{Based on observations made with the NASA/ESA Hubble Space
Telescope, obtained from MAST at the Space Telescope Science Institute, which is
operated by the Association of Universities for Research in Astronomy, Inc.,
under NASA contract NAS 5-26555. These observations are associated with
program \#13859.}

\author{L. Fossati\altaffilmark{2}}
\affil{Space Research Institute, Austrian Academy of Sciences, Schmiedlstrasse 		6, A-8042 Graz, Austria}
\email{luca.fossati@oeaw.ac.at}
\and
\author{K. France\altaffilmark{3}}
\affil{Laboratory for Atmospheric and Space Physics, University of Colorado, 600 UCB, Boulder, CO 80309, USA}
\email{kevin.france@colorado.edu}
\and
\author{T. Koskinen}
\affil{Lunar and Planetary Laboratory, University of Arizona, 1629 East University Boulevard, Tucson, AZ 85721-0092, USA}
\email{tommi@lpl.arizona.edu}
\and
\author{I.~G. Juvan}
\affil{Space Research Institute, Austrian Academy of Sciences, Schmiedlstrasse 		6, A-8042 Graz, Austria}
\email{ines.juvan@oeaw.ac.at}
\and
\author{C.~A. Haswell}
\affil{Department of Physical Sciences, Open University, Walton Hall, Milton Keynes MK7 6AA, UK}
\email{C.A.Haswell@open.ac.uk}
\and
\author{M. Lendl}
\affil{Space Research Institute, Austrian Academy of Sciences, Schmiedlstrasse 		6, A-8042 Graz, Austria}
\email{monika.lendl@oeaw.ac.at}

\altaffiltext{2}{Argelander-Institut f\"ur Astronomie der Universit\"at Bonn, 			Auf dem H\"ugel 71, 53121, Bonn, Germany}
\altaffiltext{3}{Center for Astrophysics and Space Astronomy, University of 			Colorado, 389 UCB, Boulder, CO 80309, USA}

\begin{abstract}
Several transiting hot Jupiters orbit relatively inactive main-sequence stars. For some of those, the \logR\ activity parameter lies below the basal level ($-$5.1). Two explanations have been proposed so far: (i) the planet affects the stellar dynamo, (ii) the \logR\ measurements are biased by extrinsic absorption, either by the interstellar medium (ISM) or by material local to the system. We present here Hubble Space Telescope/COS far-UV spectra of WASP-13, which hosts an inflated hot Jupiter and has a measured \logR\ value ($-$5.26), well below the basal level. From the star's spectral energy distribution we obtain an extinction \ebv\,=\,0.045$\pm$0.025\,mag and a distance $d$\,=\,232$\pm$8\,pc. We detect at $\gtrsim$4$\sigma$ lines belonging to three different ionization states of carbon (\ion{C}{1}, \ion{C}{2}, and \ion{C}{4}) and the \ion{Si}{4} doublet at $\sim$3$\sigma$. Using far-UV spectra of nearby early G-type stars of known age, we derive a \ion{C}{4}/\ion{C}{1} flux ratio--age relation, from which we estimate WASP-13's age to be 5.1$\pm$2.0\,Gyr. We rescale the solar irradiance reference spectrum to match the flux of the \ion{C}{4} 1548 doublet. By integrating the rescaled solar spectrum, we obtain an XUV flux at 1\,AU of 5.4\,\ergscm. We use a detailed model of the planet's upper atmosphere, deriving a mass-loss rate of 1.5$\times$10$^{11}$\,g\,s$^{-1}$. Despite the low \logR\ value, the star shows a far-UV spectrum typical of middle-aged solar-type stars, pointing toward the presence of significant extrinsic absorption. The analysis of a high-resolution spectrum of the \ion{Ca}{2}\,H\&K lines indicates that the ISM absorption could be the origin of the low \logR\ value. Nevertheless, the large uncertainty in the \ion{Ca}{2} ISM abundance does not allow us to firmly exclude the presence of circumstellar gas.
\end{abstract}
%



\section{Introduction}\label{sec:introduction}
To date, over 200 close-in ($<$0.1\,AU) giant planets (hot Jupiters) have been discovered. These planets were not expected to survive for long owing to the rapid evaporation of their atmospheres under the effect of the strong stellar irradiation. Hubble Space Telescope (HST) ultraviolet (UV) observations confirmed that the atmospheres of hot Jupiters such as HD\,209458\,b, HD\,189733\,b, and WASP-12\,b are evaporating \citep{vidal2003,linsky2010,haswell2012,bourrier2013,fossati2015}. The stellar extreme-UV (EUV) flux heats the atmosphere of hot Jupiters to
temperatures up to 10$^4$\,K, leading to atmospheric mass loss on the order of 10$^9$\,$\mathrm{g\,s}^{-1}$ \citep{lecavelier2007,ehrenreich2011,koskinen2013a,koskinen2013b}, which, however, does not seem to significantly reduce the planetary mass within the star's main-sequence lifetime \citep{yelle2004,garcia2007}.

A number of hot Jupiters are found orbiting around low-activity stars (\logR\,$<$\,-5.0), which is loosely correlated with the biases that stellar activity introduces in detecting planets around active stars. Stellar activity, directly related to the EUV flux \citep[e.g.,][]{piters1997}, plays a crucial role in shaping planets' structure and evolution \citep[e.g.,][]{hubeny2003,fortney2008,davis2009,zahnle2009,howard2012,hu2012,mordasini2012,beauge2013}. It is therefore crucial to carefully study the activity of fairly inactive stars to accurately infer their EUV fluxes.

Given their intrinsically low levels of chromospheric emission, all but the nearest relatively inactive solar-type stars have proved difficult to study at UV wavelengths. However, given the new interest in exoplanet host stars and technological advances in spectroscopic instrumentation on HST in the preceding decade \citep[e.g., the installation of the Cosmic Origins Spectrograph;][]{green2012}, there has been a renewed interest in the study of relatively inactive, late-type dwarf stars \citep[e.g.,][]{france2013,linsky2013,stelzer2013}. The lack of information about the high-energy radiation of inactive stars has limited reliable calculations of atmosphere and evolution models of planets orbiting close to such stars. \citet{linsky2013,linsky2014} used nearby stars, for which optical, far-UV (FUV), and EUV spectra are available, to construct relations between EUV fluxes and various UV and optical chromospheric emission features. These relations have typical uncertainties of the order of 15--50\%, depending on the considered wavelength region. Therefore, more detailed observations are needed to improve the EUV flux estimates for the most inactive stars. Some hot-Jupiter hosts, such as WASP-12, have an anomalously low activity level, well below the range covered by the reference stars adopted by \citet{linsky2013,linsky2014}.

The late F-type star WASP-12 hosts one of the hottest and most bloated known exoplanets \citep{hebb2009}. Near-UV (NUV) HST COS data revealed that the upper atmosphere of WASP-12\,b overfills its Roche lobe and is losing mass rapidly \citep{fossati2010a,haswell2012}. The HST data also revealed the presence of a broad depression in place of the normally bright emission cores in the \ion{Mg}{2}\,h\&k resonance lines \citep{haswell2012}. \citet{fossati2013} showed that the \ion{Ca}{2}\,H\&K line cores also present broad and deep depressions, similar to the \ion{Mg}{2}\,h\&k line profiles. The anomaly is always present, regardless of the planet's orbital phase. The absence of \ion{Mg}{2} core emission was unexpected given the star's spectral type and age \citep[F9V, effective temperature of 6250\,K, age\,$<$\,2.65\,Gyr;][]{fossati2010b}. This peculiarity is reflected in an anomalously low \logR\ value\footnote{The \logR\ value is a measure of the chromospheric emission in the core of the \ion{Ca}{2}\,H\&K resonance lines \citep{duncan1991}. For reference, the average solar \logR\ value is $-$4.902$\pm$0.063 \citep[95\% confidence level;][]{mamajek2008}. At maximum of activity (e.g., on cycle 22) the solar \logR\ value is typically about $-$4.780, while at minimum the solar \logR\ value is typically about $-$4.966. The Sun during the Maunder minimum was believed to have a \logR\ value of $-$5.102 \citep{baliunas1990,donahue1998,keil1998,radick1998,livingston2007}.} of -5.500 \citep{knutson2010}, far below the basal flux level of main-sequence late-type stars of \logR\,=\,-5.1 \citep{wright2004}.

\citet{fossati2013} analyzed the origin of the anomalous line cores, concluding that extrinsic absorption by material local to the WASP-12 system is the most likely cause: gas escaping from the heavily irradiated planet could form a stable and translucent circumstellar cloud \citep{haswell2012}. Theoretical support to this idea has been given by \citet{cohen2011}, \citet{lanza2014}, \citet{matsakos2015}, and \citet{fossati2015b}. The H$_2$ fluorescence observed around planet-hosting M-type stars could also be a similar manifestation \citep{france2013}. Nevertheless, detailed hydrodynamic simulations and radiative transfer calculations should be carried out to explore the spreading of the specific ions escaping from the planet and whether their column density would be large enough to affect the stellar core emission in the \ion{Mg}{2}\,h\&k and \ion{Ca}{2}\,H\&K resonance lines.

The anomalously low stellar activity index of WASP-12 may therefore be a direct consequence of the extra line core absorption. For stars hosting close-in evaporating planets, it may not be therefore possible to evaluate the actual stellar activity from the \logR\ value. \citet{pillitteri2014} explained the low activity level and X-ray flux of WASP-18, an F9V star hosting a massive hot Jupiter, by involving an anomalous decrease of the stellar activity caused by the planet affecting the stellar dynamo. Because of the high formation temperature, several lines in the FUV would not be absorbed by the interstellar medium (ISM) and circumstellar cloud, allowing one to retrieve the actual stellar activity, which can then be compared to that obtained from other indicators. The FUV flux level discriminates therefore between the two explanations for the depressed \logR\ values observed for several hot-Jupiter host stars.

We present here COS FUV observations of the G1V star WASP-13 \citep[$M$\,=\,1.187\,\Mo\ and $R$\,=\,1.574\,\Ro;][]{yilen2013}, which hosts an extremely inflated hot Jupiter \citep[$M_{\rm p}$\,=\,0.500$\pm$0.037\,$M_{\rm J}$ and $R_{\rm p}$\,=\,1.407$\pm$0.052\,$R_{\rm J}$;][]{skillen2009,yilen2013} and displays an activity well below the basal level with \logR\,=\,-5.263$\pm$0.09 \citep{knutson2010,fossati2013}, where the uncertainty is the typical standard deviation in the \logR\ value caused by variability within the activity cycle of an inactive star \citep{lanza2014}. There is a rather large spread of ages for WASP-13 present in the literature: 8.5$\pm$5.0\,Gyr \citep{skillen2009}, 7.4$\pm$0.4\,Gyr \citep{barros2012}, 4--5.5\,Gyr \citep{yilen2013}, and 12.59$\pm$0.01\,Gyr \citep{bonfanti2015}.

Figures 3 and 4 of \citet{wright2004} indicate that there are stars with a \logR\ value below the basal level, but these stars are subgiants and hence evolved off the main sequence. Even without recognizing the distinction between main-sequence and evolved stars in terms of activity, the \logR\ index of WASP-13 is among the lowest in the literature for solar-type stars. It is therefore worth examining whether the \logR\ index is a good measure of the ``true'' activity of the star; this is our primary aim. We use higher-ionization FUV emission lines, not affected by extrinsic absorption (e.g., \ion{C}{4} and \ion{Si}{4}), to estimate the true stellar activity, derive the EUV flux, and attempt to identify the origin of the low \logR\ value.

Section~\ref{sec:observations} describes the HST observations and the applied data reduction. In Section~\ref{sec:results} we gather the results: analysis of the FUV planet transit (Section~\ref{sec:transit}), estimate of the distance to the star and interstellar extinction (Section~\ref{sec:sed}), and measurements of the FUV stellar emission lines (Section~\ref{sec:line_flux}). In Section~\ref{sec:discussion} we estimate the stellar chromospheric activity (Section~\ref{sec:activity}), age (Section~\ref{sec:activity}), and high-energy flux (Section~\ref{sec:xuv}), and perform an analysis aiming at the detection of the circumstellar material (Section~\ref{sec:cloud}). We draw the conclusions in Section~\ref{sec:conclusions}.
\section{Observations and data reduction}\label{sec:observations}
We observed the planet-hosting star WASP-13 using the FUV channel of HST/COS. The observations were conducted on 2015 April 24 along four consecutive HST orbits, timed with the planet's primary transit. The observing log is given in Table~\ref{tab:obs.log}. Each spectrum was taken using the FUV G140L grating at the 1105\,\AA\ setting, which provides a continuous spectrum over the 1128--2236\,\AA\ wavelength range at a spectral resolution of $R\sim$3000, in TIME-TAG mode. To achieve the maximum possible stability to attempt the detection of the planet transit, we adopted one single FP position (i.e., FP-POS\,=\,3; see the COS user manual\footnote{\tt http://www.stsci.edu/hst/cos/documents/handbooks/current/cos\_cover.html} for details). 
\begin{table}[ht]
\caption[ ]{Observing Log of the COS observations of WASP-13. Columns (2) and (4) give respectively the Julian date (JD$-$2,450,000) and the range of planet orbital phases covered by each exposure, respectively. Column (3) gives the total exposure time in seconds. The exposure time adopted for the first orbit is shorter than that of the following orbits, because of the HST target pointing and acquisition preceeding the science exposure.}
\label{tab:obs.log}
\begin{center}
\begin{tabular}{l|ccc}
\hline
\hline
 & JD           & Exp. Time & Pl. Orbital \\
 & $-$2,450,000 & (s)       & Phase Range \\
\hline
Orbit 1 & 7133.788574 & 2522.176 & 0.973\,$-$\,0.980 \\
Orbit 2 & 7133.851562 & 2971.168 & 0.987\,$-$\,0.995 \\
Orbit 3 & 7133.917969 & 2971.200 & 1.002\,$-$\,1.010 \\
Orbit 4 & 7133.983887 & 2971.200 & 1.017\,$-$\,1.025 \\
\hline
\end{tabular}
\end{center}
\end{table}

\begin{figure}
\begin{center}
\includegraphics[width=\hsize,clip]{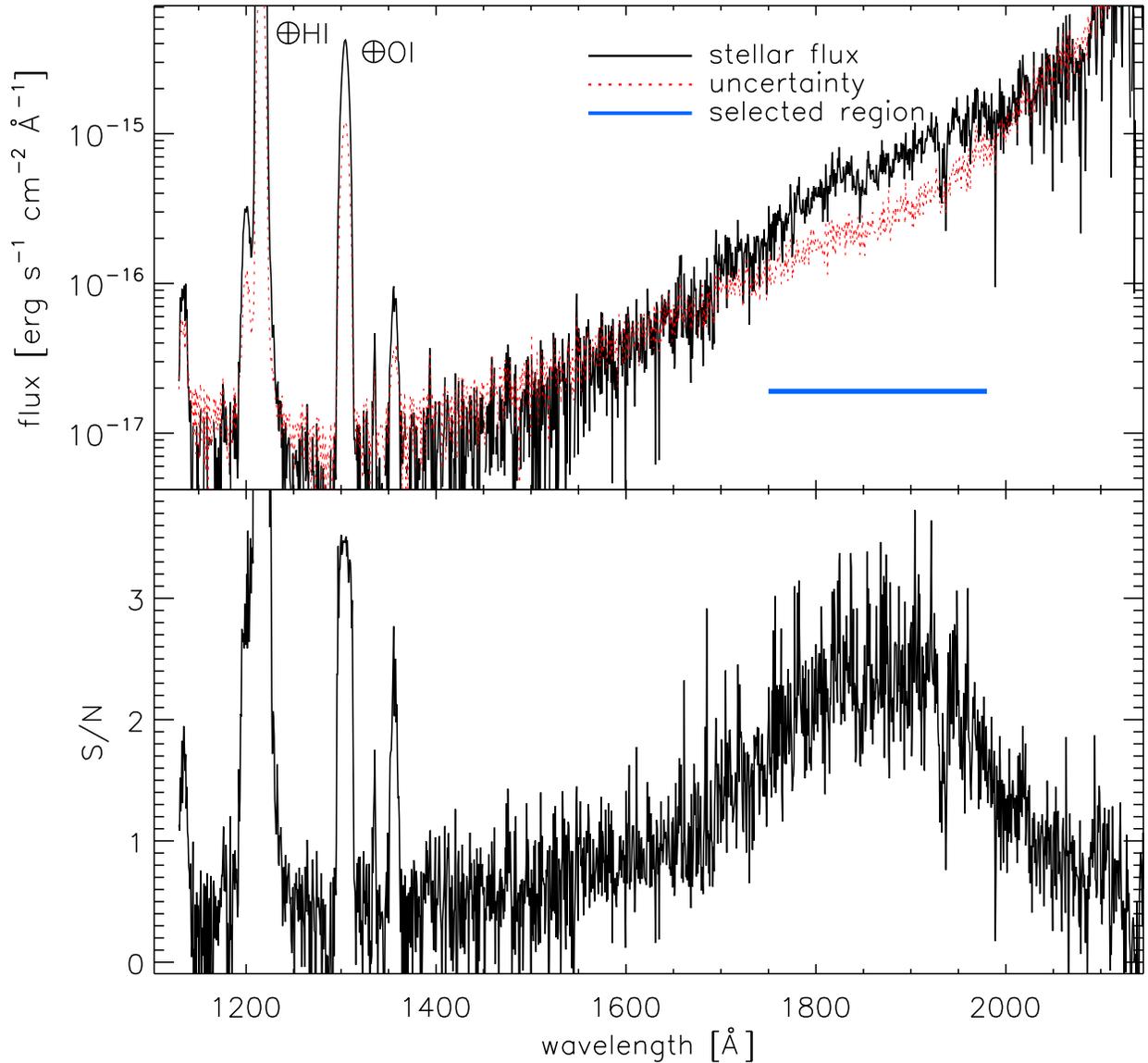}
\caption{Top: flux-calibrated co-added COS spectrum of WASP-13 (black solid line) and corresponding uncertainties (red dotted line). The strong emission lines at short wavelengths are caused by \ion{H}{1} and \ion{O}{1} geocoronal emission. The blue horizontal bar between 1750 and 1980\,\AA\ indicates the spectral region used to attempt the detection of the planet transit (see Section~\ref{sec:transit}). Bottom: S/N of the co-added COS spectrum as a function of wavelength.}
\label{fig:whole-spectrum}
\end{center}
\end{figure}

We retrieved the data, reduced with CALCOS V.3.0, from the MAST\footnote{\tt http://archive.stsci.edu/} archive and co-added the four flux-calibrated spectra, using the routine described by \citet{danforth2010}, in order to obtain one spectrum with the highest possible signal-to-noise ratio (S/N). The final co-added FUV spectrum of WASP-13 is shown in Figure~\ref{fig:whole-spectrum}. The strong emission lines at short wavelengths are caused by geocoronal emission from atomic hydrogen and oxygen, while the rise in flux at longer wavelengths is due to the emission from the stellar photosphere, indicating that the star was indeed detected in the FUV.
%
\section{Results}\label{sec:results}
\subsection{Planet transit}\label{sec:transit}
To investigate the FUV transit properties of WASP-13\,b, we adopted the ephemeris given by \citet{barros2012}, which results in having the second and third exposures being obtained during the transit, while the first/fourth exposures are obtained before/after the ingress/egress. Given the low stellar flux present in the emission lines (see Sect~\ref{sec:line_flux}), we attempted the detection of the planet transit in the 1750--1980\,\AA\ wavelength range, dominated by photospheric flux; this is the wavelength range that minimizes the uncertainties on the integrated fluxes (see Figure~\ref{fig:whole-spectrum}).

\begin{figure}
\begin{center}
\includegraphics[width=\hsize,clip]{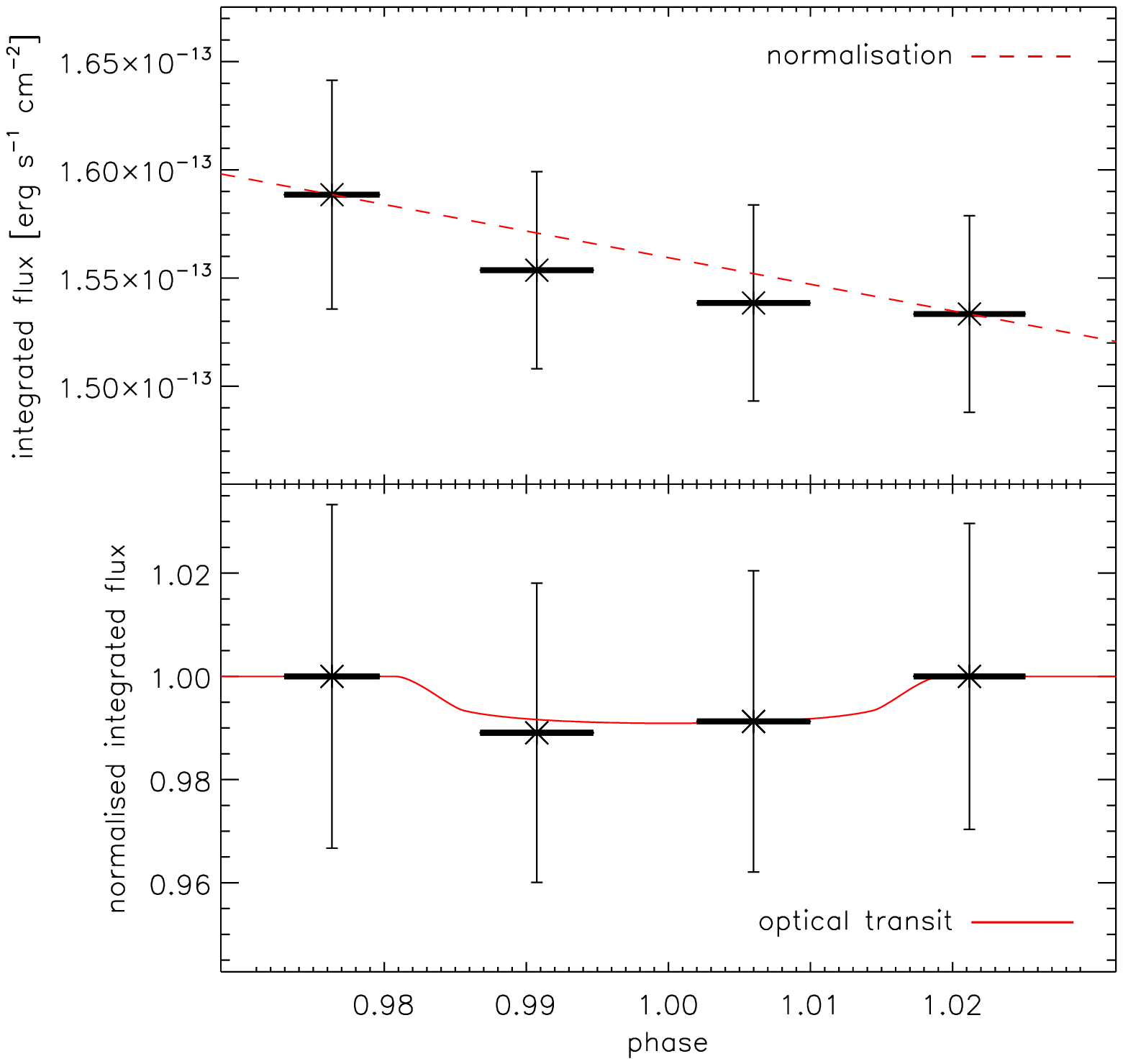}
\caption{Top panel: flux obtained from each spectrum integrating within the 1750--1980\,\AA\ wavelength range as a function of the planet orbital phase. The thick horizontal bars indicate the phase range covered by the observations. The red dashed line shows the adopted out-of-transit (i.e., first and last HST orbit) normalization. Bottom panel: same as top panel, but with the normalized fluxes. The red solid line shows a model of the optical transit in the $r'$ band.}
\label{fig:transit}
\end{center}
\end{figure}

Figure~\ref{fig:transit} shows the results of the integration in wavelength for each of the four spectra. The uncertainties are consistent with the scatter obtained by splitting each point into three observations of equal exposure time using the TIME-TAG. To compare the observations with the optical transit, calculated in the $r'$ band using the ephemeris and planet parameters given by \citet{barros2012}, we divided each point by the line connecting the out-of-transit integrated fluxes. Despite the large considered wavelength region, the photometric uncertainties, of the order of 3.5\%, are too large to detect the planet transit, which has a depth of about 1\% (in flux). The planet radius observed in the 1750--1980\,\AA\ wavelength range is therefore likely to be smaller than 2.9\,$R_{\rm Jup}$.

We then calculated the Rayleigh scattering transit depth \citep[e.g.,][]{hubbard01} to check whether the nondetection of the transit in the FUV would give any constraints on the presence, or not, of Rayleigh scattering in the planet's atmosphere. We adopted the system parameters given by \citet{barros2012} and assumed a temperature of 1500\,K at pressures higher than 10$^{-6}$\,bar. We calculated the equilibrium mixing ratios of H$_2$ and H as a function of altitude based on the temperature and pressure and used Rayleigh scattering cross sections from \citet{dalgarno65} for H$_2$. We found that Rayleigh scattering by H \citep{lee04}, which is also included in our calculation, is negligible for WASP-13\,b. Our results indicate that the transit depth due to Rayleigh scattering by H$_2$ at 1750--1980\,\AA\ is only about 9\% deeper than in the optical. The photometric uncertainties are therefore too large to provide significant ($>$2$\sigma$) constraints on the presence of Rayleigh scattering in the planet's atmosphere.  
\subsection{Distance to the star and interstellar extinction}\label{sec:sed}
To estimate the interstellar extinction \ebv\ along the WASP-13's line of sight, we used the available infrared, optical, and UV photometry, together with synthetic stellar fluxes, to fit the star's spectral energy distribution (SED). To construct the observed SED, we considered Two Micron All Sky Survey (2MASS) and WISE infrared photometry from \citet{cutri2013} with calibrations from \citet{bliek1996} and \citet{wright2010}, respectively, TYCHO photometry from \citet{hog2000} with calibrations from \citet{mann2015}, Johnson photometry from \citet[][$B$-band]{kharchenko2001} and \citet[][$V$- and $I$-band]{droege2006} with calibrations from \citet{bessel1998}, and GALEX UV fluxes from \citet{shkolnik2013}.

We calculated the stellar synthetic fluxes with the model atmosphere code \llm\ \citep{shulyak2004}, using the atmospheric parameters given by \citet{yilen2013}: effective temperature (\Teff) of 5989\,K, surface gravity (\logg) of 4.16, and solar metallicity \citep{asplund2009}. Despite the noninclusion of molecular opacities in \llm, a comparison with {\sc marcs} models \citep{marcs} shows that \llm\ can be safely used for stars at least as cool as the Sun, hence WASP-13 as well. When fitting synthetic fluxes to observed fluxes, the distance to the star and the stellar radius are degenerate. We adopted the stellar radius of $R_{\star}$\,=\,1.574$\pm$0.048\,\Ro\ \citep{yilen2013} to derive the distance to the star and the interstellar extinction, the latter modeled using the \citet{fitzpatrick1999} extinction model and assuming a total-to-selective extinction coefficient of $R_{\rm V}$\,=\,3.1.

Since \llm\ calculates solely the photospheric fluxes and does not account for chromospheric/coronal emission, we only considered the NUV band of the GALEX fluxes in the fit. Here the major chromospheric contribution would come from the \ion{Mg}{2}\,h\&k chromospheric line core emission, which for WASP-13 is negligible compared to the total NUV photospheric emission (see Section~\ref{sec:xuv}). During the SED analysis, we noticed that the TYCHO $V$- and Johnson $I$-band fluxes were always far higher than the synthetic fluxes best fitting all other constraints; hence, we decided to exclude them from the fit. We have not been able to find the origin of these discrepancies. We also did not consider the reddest WISE flux in the fit, because it is an upper limit \citep{cutri2013}. The fit was done taking into account the transmission curves specific for each filter, downloaded from the Virtual Observatory\footnote{{\tt http://svo2.cab.inta-csic.es/svo/theory/fps3/}}.

\begin{figure}
\begin{center}
\includegraphics[width=14cm,clip]{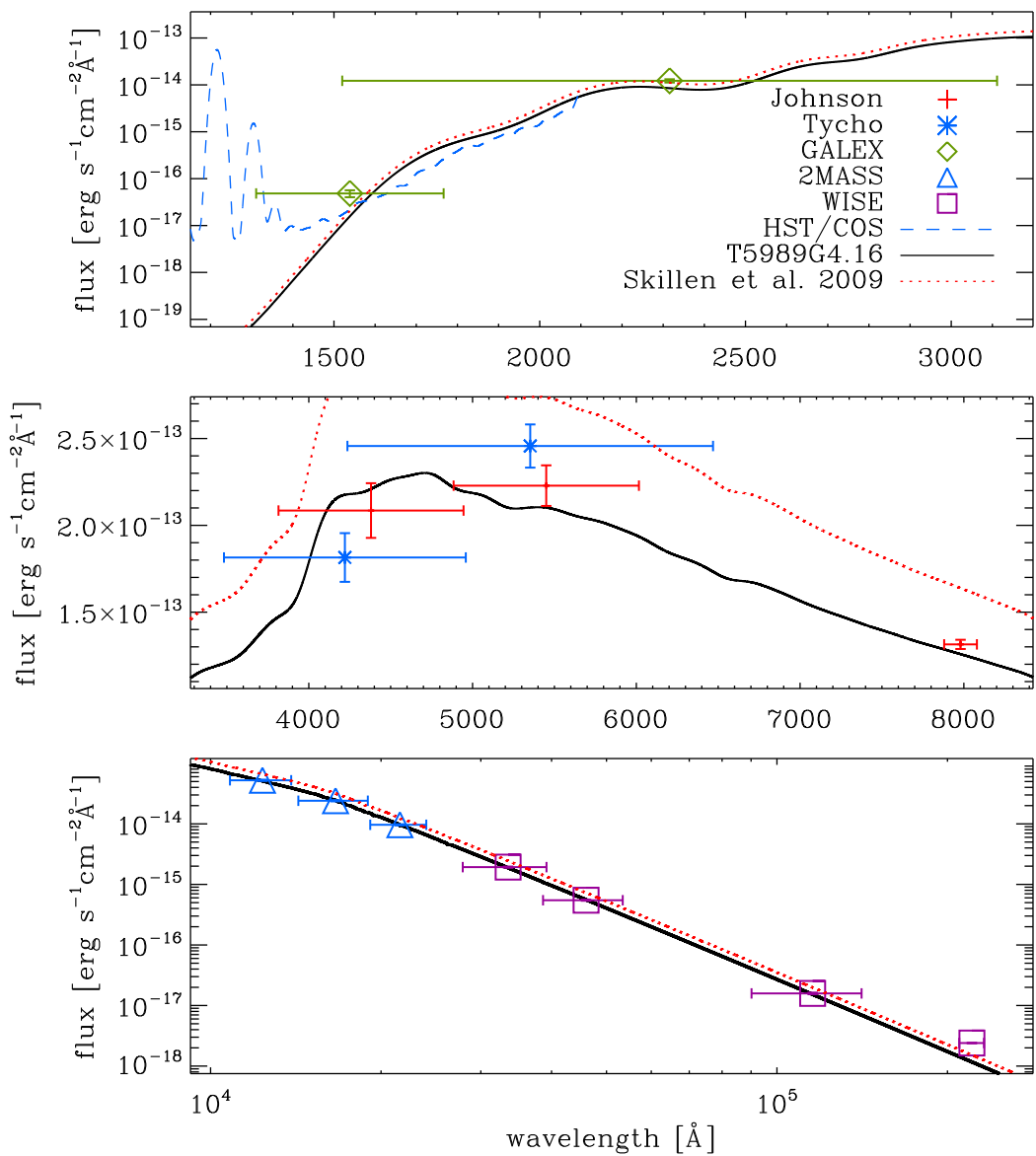}
\caption{Comparison between \llm\ synthetic fluxes (full black line), calculated with the atmospheric parameters given by \citet{yilen2013}, with COS (dashed blue line) and GALEX fluxes (green rhombi), and with Johnson (red crosses), TYCHO (blue asterisks), 2MASS (blue triangles), and WISE (purple squares) photometry converted to physical units. The horizontal lines indicate the wavelength range covered by each photometric point. For visualization purposes, the synthetic and COS fluxes have been convolved with a Gaussian having a half-width at half-maximum of 100 and 10\,\AA, respectively. The red dotted line shows the SED of WASP-13 obtained considering the stellar parameters and distance given by \citet{skillen2009} and an extinction of \ebv\,=\,0.045$\pm$0.025\,mag.}
\label{fig:sed}
\end{center}
\end{figure}

The best fit to the observed SED is shown in Figure~\ref{fig:sed} and is obtained with a distance to the star of $d$\,=\,232$\pm$8\,pc and an interstellar extinction of \ebv\,=\,0.045$\pm$0.025\,mag, with a $\chi^2$ of 13.5 and 8 degrees of freedom. These uncertainties incorporate the error bar on the stellar radius. In Figure~\ref{fig:sed} we plot also the observed COS fluxes, but do not consider them for the fit, though the photospheric-dominated part agrees well with the best-fitting synthetic fluxes.

The derived distance is in clear disagreement with that given in the WASP-13\,b discovery paper $d$\,=\,155$\pm$18\,pc \citep{skillen2009} and currently listed in the available exoplanet databases (e.g., {\tt http://exoplanets.org/} and {\tt http://exoplanet.eu/}). We have confidence that our result is correct, as we explain now. Ignoring extinction, it is possible to analytically estimate the distance based on the stellar apparent magnitude and luminosity:
\begin{equation}
m_{\star}=m_{\odot}-2.5\,\mathrm{log}{\left [ \frac{L_{\star}}{L_{\odot}} \left (\frac{d_{\odot}}{d_{\star}} \right )^2 \right ] }\,\,,
\end{equation} 
where $m_{\star}$ is the apparent stellar Johnson $V$-band magnitude, $m_{\odot}$ is the apparent Johnson $V$-band magnitude of the Sun (we adopted $m_{\odot}$\,=\,$-$26.73\,mag), $L_{\star}$ is the stellar luminosity (calculated from \Teff\ and radius), $L_{\odot}$ is the solar luminosity, $d_{\odot}$ is the distance to the Sun (1\,AU), and $d_{\star}$ is the distance to the star. In this way we obtained a distance $d_{\star}$\,=\,233\,pc, which is in excellent agreement with that obtained from our SED fitting. We further compared the derived extinction with the value given by the Galactic extinction maps of \citet{amores2005}, assuming a distance of 232\,pc, obtaining \ebv\,=\,0.050\,mag, which is in good agreement with our result. Figure~\ref{fig:sed} shows also that the SED obtained considering the stellar parameters and distance given by \citet{skillen2009} does not fit the available photometry.

We used the \ebv\ value obtained from the SED fitting to derive the ISM \ion{H}{1}, \ion{C}{1}, \ion{C}{2}, \ion{C}{4}, \ion{Si}{4}, \ion{Mg}{2}, \ion{Ca}{2}, and \ion{Na}{1} column density. We considered solar abundances from \citet{asplund2009} and the ISM ionization mix by \citet[][their model 2 - Table~5]{frisch2003}. We derived the column densities considering four different $N_{\rm HI}$--\ebv\ conversions\footnote{Given the small \ebv\ value, we assumed that hydrogen is not in molecular form \citep[e.g.,][]{rachford2002}.} given by \citet{savage1979}, \citet{diplas1994}, \citet{guver2009}, and \citet{gudennavar2012}. For \ion{Mg}{2}, \ion{Ca}{2}, and \ion{Na}{1} we also derived the column densities by considering the ISM ion abundance as a function of the \ion{H}{1} column density given by \citet{wakker2000}. The results are listed in Table~\ref{tab:ebv-columnDensities}.
\begin{table*}[ht]
\caption[ ]{ISM column densities (in cm$^{-2}$) derived for \ion{H}{1}, \ion{C}{1}, \ion{C}{2}, \ion{C}{4}, \ion{Si}{4}, \ion{Mg}{2}, \ion{Ca}{2}, and \ion{Na}{1} on the basis of the $N_{\rm HI}$--\ebv\ conversions given by \citet[][SM79]{savage1979}, \citet[][DS94]{diplas1994}, \citet[][GO09]{guver2009}, and \citet[][G12]{gudennavar2012}. The column densities are given using either the ion ISM abundances of \citet[][WM00]{wakker2000} or the solar element abundances by \citet[][A09]{asplund2009} in combination with the ISM ionisation mix by \citet[][FS03]{frisch2003}. The uncertainties account for the error bar on \ebv\ of 0.025\,mag, only.}
\label{tab:ebv-columnDensities}
\hspace{-1.7cm}
\begin{tabular}{l|cccccccc}
\hline
\hline
Assumptions & log\,N$_{\rm HI}$ & log\,N$_{\rm CI}$ & log\,N$_{\rm CII}$ & log\,N$_{\rm CIV}$ & log\,N$_{\rm SiIV}$ & log\,N$_{\rm MgII}$ & log\,N$_{\rm CaII}$ & log\,N$_{\rm NaI}$ \\
\hline
SM79+A09+FS03 & 20.42$^{+0.19}_{-0.36}$ & 13.47$^{+0.19}_{-0.36}$ & 16.79$^{+0.19}_{-0.36}$ & 0.00$^{+0.00}_{-0.00}$ & 11.39$^{+0.19}_{-0.36}$ & 15.89$^{+0.19}_{-0.36}$ & 12.97$^{+0.19}_{-0.36}$ & 11.77$^{+0.19}_{-0.36}$ \\
DS94+A09+FS03 & 20.35$^{+0.19}_{-0.36}$ & 13.40$^{+0.19}_{-0.36}$ & 16.72$^{+0.19}_{-0.36}$ & 0.00$^{+0.00}_{-0.00}$ & 11.32$^{+0.19}_{-0.36}$ & 15.82$^{+0.19}_{-0.36}$ & 12.90$^{+0.19}_{-0.36}$ & 11.70$^{+0.19}_{-0.36}$ \\
GO09+A09+FS03 & 20.49$^{+0.19}_{-0.36}$ & 13.54$^{+0.19}_{-0.36}$ & 16.87$^{+0.19}_{-0.36}$ & 0.00$^{+0.00}_{-0.00}$ & 11.46$^{+0.19}_{-0.36}$ & 15.97$^{+0.19}_{-0.36}$ & 13.04$^{+0.19}_{-0.36}$ & 11.84$^{+0.19}_{-0.36}$ \\
G12+A09+FS03  & 20.81$^{+0.05}_{-0.05}$ & 13.86$^{+0.05}_{-0.05}$ & 17.19$^{+0.05}_{-0.05}$ & 0.00$^{+0.00}_{-0.00}$ & 11.79$^{+0.05}_{-0.05}$ & 16.29$^{+0.05}_{-0.05}$ & 13.36$^{+0.05}_{-0.05}$ & 12.17$^{+0.05}_{-0.05}$ \\
\hline
SM79+WM00 & 20.42$^{+0.19}_{-0.36}$ & -- & -- & -- & -- & 14.97$^{+0.14}_{-0.27}$ & 11.94$^{+0.04}_{-0.08}$ & 12.15$^{+0.16}_{-0.30}$ \\
DS94+WM00 & 20.35$^{+0.19}_{-0.36}$ & -- & -- & -- & -- & 14.91$^{+0.14}_{-0.27}$ & 11.93$^{+0.04}_{-0.08}$ & 12.09$^{+0.16}_{-0.30}$ \\
GO09+WM00 & 20.49$^{+0.19}_{-0.36}$ & -- & -- & -- & -- & 15.02$^{+0.15}_{-0.27}$ & 11.96$^{+0.04}_{-0.08}$ & 12.21$^{+0.16}_{-0.30}$ \\
G12+WM00  & 20.81$^{+0.05}_{-0.05}$ & -- & -- & -- & -- & 15.27$^{+0.03}_{-0.04}$ & 12.03$^{+0.01}_{-0.01}$ & 12.48$^{+0.04}_{-0.04}$ \\
\hline
\end{tabular}
\end{table*}


For the elements considered here, \ion{C}{2}, \ion{Si}{2}, \ion{Mg}{2}, \ion{Ca}{2}, and \ion{Na}{1} are the main ionization stages in the ISM. At the low \ebv\ value of WASP-13, almost all of the carbon in the diffuse ISM is in the singly ionized state \citep{snow2006}. \ion{C}{4} is not typically observed in the local ISM, which is why Table~\ref{tab:ebv-columnDensities} lists a null column density \citep{frisch2003}, though a detection of \ion{C}{4} in the local ISM has been reported by \citet{welsh2010}. Taking into account the uncertainty on \ebv, the three different conversions to the log\,$N_{\rm HI}$ column density from \ebv\ give a value of log\,$N_{\rm HI}$ in the $\sim$20.0--20.8\,dex range. This spread of about 0.8\,dex is then also present in the column densities derived for the other ions. The use of the ISM ion abundances of \citet{wakker2000} leads to column densities that are about 1\,dex smaller for \ion{Mg}{2} and \ion{Ca}{2} and about 0.4\,dex larger for \ion{Na}{1}, compared to what was obtained assuming solar abundances in combination with the ISM ionization mix by \citet{frisch2003}. 

\citet{frisch2003} showed that the average chemical composition of the local ISM (i.e., within 200\,pc) is about 60\%--70\% subsolar, and \citet{savage1996}, looking at high Galactic latitude halo stars, arrived at the same conclusion. We therefore believe that the values reported in the lower half of Table~\ref{tab:ebv-columnDensities}, obtained using the ISM ion abundances of \citet{wakker2000}, should be more realistic than those based on solar abundances and hence adopt those hereafter.
\subsection{Intrinsic line emission}\label{sec:line_flux}
Despite the rather large distance to the star (see Section~\ref{sec:sed}) and the expected low intrinsic stellar activity, a handful of stellar emission lines are visible in the FUV spectrum. We measured an integrated flux for the \ion{C}{1} multiplets at $\sim$1560 and $\sim$1657\,\AA, the \ion{C}{2} doublet at $\sim$1335\,\AA, the \ion{C}{4} doublet at $\sim$1548\,\AA, and the \ion{Si}{4} doublet at $\sim$1394 and $\sim$1403\,\AA. No further stellar lines appear to be strong enough to be visible in the spectrum.

To measure the fluxes, we integrated over a wavelength range chosen to cover all lines within each given feature (hereafter called ``line'' regions), taking into account the spectral resolution. We did this with the guidance of an HST/Space Telescope Imaging Spectrograph (STIS) high-resolution spectrum of $\alpha$\,Cen\,A, downloaded from the StarCAT\footnote{{\tt http://casa.colorado.edu/$\sim$ayres/StarCAT/}} catalog \citep{ayres2010} and convolved to the resolution of the COS spectrum. 

\begin{figure*}
\begin{center}
\vspace{-1.0cm}
\includegraphics[width=15.0cm,clip]{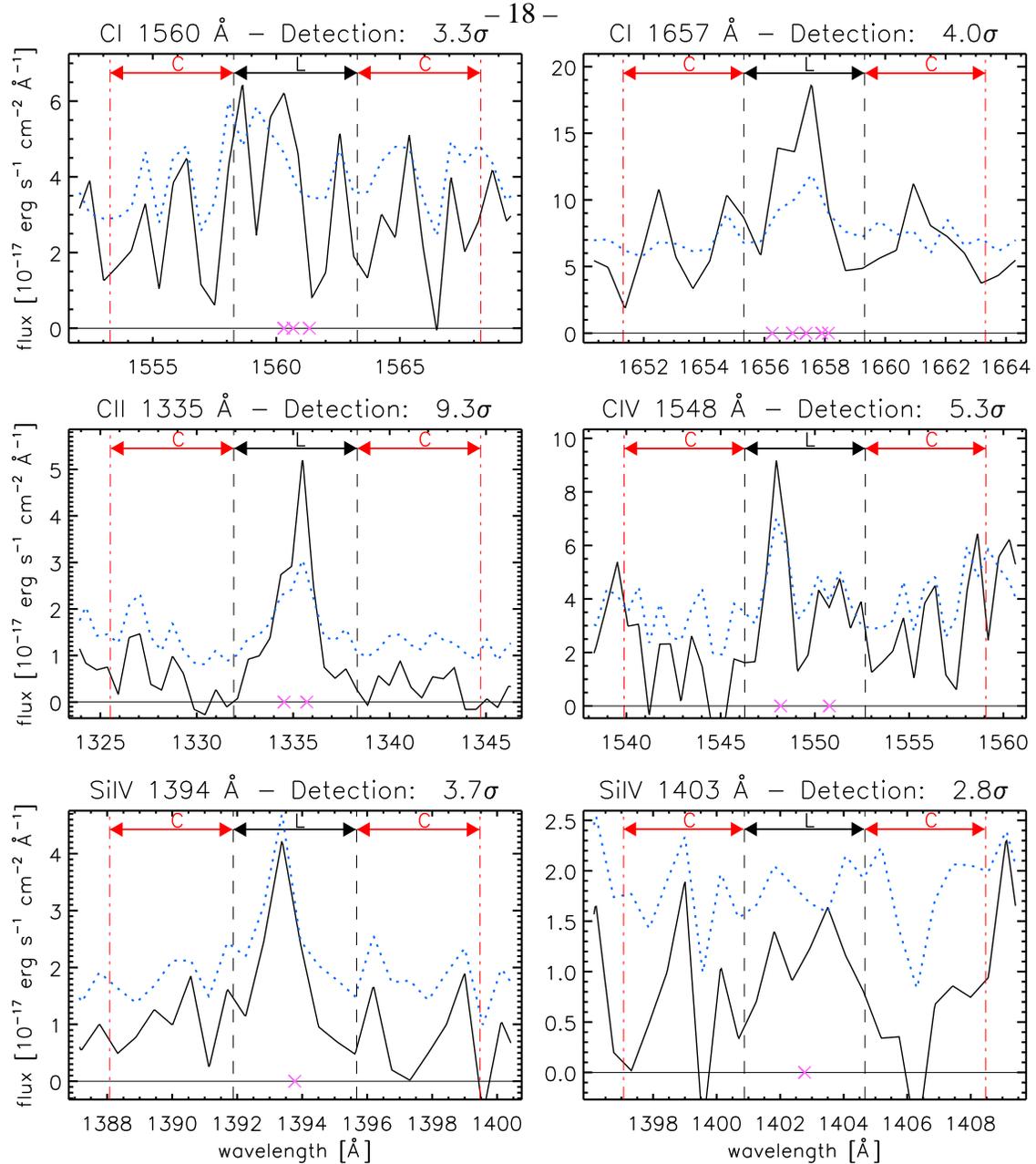}
\caption{COS spectrum of WASP-13 in the region of the six analyzed features. The black solid line and the blue dashed line show, respectively, the stellar flux and its uncertainty in units of 10$^{-17}$ erg\,s$^{-1}$\,cm$^{-2}$\,\AA$^{-1}$. The crosses on the bottom of each panel indicate the expected position of stellar emission lines for the transition given above each panel. The vertical black dashed lines mark the regions used to measure the integrated line fluxes, further indicated by black arrows and an L on the top of each panel. The red dot-dashed lines mark the two regions for each feature used to derive the integrated continuum flux, further indicated by red arrows and a C on the top of each panel. The $\sigma$ detection for each feature is given above each panel.}
\label{fig:lines_w13}
\end{center}
\end{figure*}

For each feature, we defined two ``continuum'' regions, one redward and one blueward of the ``line'' region, with the three regions having the same width. The continuum flux for each feature was then set as the average of the integrated fluxes calculated from the two ``continuum'' regions. The uncertainty on the continuum flux was derived by adding the uncertainties, obtained from the two continuum regions, in quadrature. We followed this procedure because the uncertainties, dominated by background subtraction, are most likely correlated; if this were not to be the case, the line detections would increase on average by about 0.5$\sigma$--1$\sigma$, depending on the considered line. We subtracted the continuum flux from the line flux to derive a continuum-subtracted line flux and its uncertainty. The wavelength regions considered for each feature are shown in Figure~\ref{fig:lines_w13}, while the resulting fluxes and line detections are listed in Table~\ref{tab:wasp13-linefluxes}.
\begin{table*}[ht]
\caption[ ]{Results obtained from the analysis of the considered FUV emission lines of WASP-13. Column (1) gives the ion and approximate wavelength. Column (2) gives the wavelength range adopted to derive the integrated line flux. Columns (3)--(5) list the integrated line, continuum, and continuum-subtracted integrated fluxes for each feature in units of 10$^{-17}$\,\ergscm. The last column lists the $\sigma$ detection.}
\label{tab:wasp13-linefluxes}
\begin{center}
\begin{tabular}{l|c|ccc|c}
\hline
\hline
Feature & Line wavelength & Line & Cont. & Cont. sub. & $\sigma$ detection \\
        & range           & flux & flux  & line flux  &                    \\
        & [\AA]           & \multicolumn{3}{c|}{[10$^{-17}$ \ergscm]}    & \\
\hline
\ion{C}{1}  - 1560\,\AA & 1558.29--1563.27 & 19.41$\pm$1.31 & 12.33$\pm$1.68 &  7.08$\pm$2.13 &  3.3 \\
\ion{C}{1}  - 1657\,\AA & 1655.35--1659.29 & 41.02$\pm$2.34 & 26.89$\pm$2.63 & 14.13$\pm$3.52 &  4.0 \\
\ion{C}{2}  - 1335\,\AA & 1331.95--1338.31 & 10.61$\pm$0.62 &  2.51$\pm$0.62 &  8.10$\pm$0.87 &  9.3 \\
\ion{C}{4}  - 1548\,\AA & 1546.31--1552.65 & 25.06$\pm$1.49 & 12.90$\pm$1.72 & 12.16$\pm$2.27 &  5.3 \\
\ion{Si}{4} - 1394\,\AA & 1391.88--1395.65 &  7.01$\pm$0.70 &  3.40$\pm$0.68 &  3.61$\pm$0.97 &  3.7 \\
\ion{Si}{4} - 1403\,\AA & 1400.88--1404.65 &  4.22$\pm$0.47 &  1.99$\pm$0.64 &  2.23$\pm$0.79 &  2.8 \\
\hline
\end{tabular}
\end{center}
\end{table*}


Table~\ref{tab:wasp13-linefluxes} shows that the \ion{C}{1} multiplet at $\sim$1657\,\AA, the \ion{C}{2} doublet at $\sim$1335\,\AA, the \ion{C}{4} doublet at $\sim$1548\,\AA, and the \ion{Si}{4} line at $\sim$1394\,\AA\ are clearly detected, while the \ion{C}{1} multiplet at $\sim$1560\,\AA\ is marginally detected at the 3.3$\sigma$ level and the \ion{Si}{4} line at $\sim$1403\,\AA\ is not detected. Three different ionization states of carbon have been detected at $\gtrsim$4$\sigma$. This result clearly indicates that despite of the measured \logR\ activity index being below the minimum threshold given for main-sequence stars, WASP-13 has a hot gas atmosphere consistent with the chromosphere and transition region lines observed in other cool main-sequence stars; WASP-13 is not completely inactive. This suggests that the low \logR\ index may be a bias caused by absorption along the WASP-13's line of sight, either by the ISM or by material local to the WASP-13 system \citep{haswell2012,fossati2013}.

In light of the ISM column densities given in Section~\ref{sec:sed}, the \ion{C}{2} doublet at $\sim$1335\,\AA\ is the most affected by ISM absorption, while the other lines, particularly \ion{C}{4} and \ion{Si}{4}, can be considered to be almost free from ISM absorption. We note that the \ion{Ca}{2} column density should be of the same order, or smaller, than that of \ion{C}{1}. In addition, in contrast to the case of the \ion{Mg}{2} lines of WASP-12 \citep{haswell2012}, the \ion{C}{2} lines of WASP-13 are not completely absorbed: for WASP-13, the presence of \ion{C}{2} emission implies that \ion{Mg}{2} line core emission should be present as well.
\section{Discussion}\label{sec:discussion}
We compare here the line flux ratios for the measured features with those of other nearby stars in order to estimate the ``true'' (i.e., free from possible biases caused by extrinsic absorptions such as ISM and/or circumstellar absorption) stellar activity level (Section~\ref{sec:activity}), age (Section~\ref{sec:activity}), and X-ray/EUV/FUV flux (Section~\ref{sec:xuv}). 
\subsection{WASP-13's chromospheric activity and age}\label{sec:activity}
To estimate the chromospheric activity of WASP-13, we compared the line flux ratios derived from the COS spectrum with those of main-sequence early G-type stars (G2--G0; similar to WASP-13) of different ages and hence activity levels. The comparison stars we selected are EK\,Dra (G1.5V; 30\,Myr), HII\,314 (G1--2V; 100\,Myr), HD\,39587 (G0V; 278\,Myr), HD\,209458 (G0V; 3\,Gyr), $\alpha$\,Cen\,A (G2V; 4.4\,Gyr), the Sun (G2V; 4.57\,Gyr), HD\,199288 (G2V; 7.7\,Gyr). These stars cover a wide range of ages; their basic parameters are listed in the first half of Table~\ref{tab:comp.stars}. The spectra of EK\,Dra, HII\,314, HD\,209458, and HD\,199288, downloaded from the MAST archive, were obtained with COS using different gratings (G130M, G160M, and G140L), while the spectra of HD\,39587 and $\alpha$\,Cen\,A, downloaded from the StarCAT catalog \citep{ayres2010}, were obtained with the STIS spectrograph and the E140M grating. Before measuring the line fluxes, all spectra, except for that of HD\,199288, obtained with COS/G140L, have been convolved with a Gaussian to match the spectral resolution of the COS spectrum of WASP-13. For the Sun we used the solar irradiance reference spectrum \citep{woods2009}.
\begin{table}[ht]
\vspace{-1cm}
\caption[ ]{Comparison stars used in Figure~\ref{fig:fit2lineratios}. The spectral types (column (2)) are taken from Simbad. Columns (3), (4), and (5) list the stars' distance, \logR\ and age (in Myr). Unless otherwise stated, the stellar distances are taken from \citet{vanLeeuwen2007} while the stars' \logR\ values and ages are taken from \citet{barnes2007}. If multiple \logR\ were available, we opted for the lowest one.}
\label{tab:comp.stars}
\begin{center}
\begin{tabular}{l|c|c|c|c}
\hline
\hline
Star & Spectral & Distance & \logR\ & Age   \\
     & type     & [pc]     &        & [Myr] \\
\hline
 EK\,Dra        & G1.5V  & 61.3$\pm$1.7    & $-$3.96$^i$ & 30$^a$    \\
 HII\,314       & G1--2V & 150$^b$         & $-$4.21$^c$ & 135$^b$   \\
 HD\,39587      & G0V    & 8.7$\pm$0.1     & $-$4.426    & 286       \\
 HD\,209458     & G0V    & 49.6$\pm$1.9    & $-$4.97$^h$ & 4000$^d$  \\
 $\alpha$Cen\,A & G2V    & 1.325$\pm$0.007 & $-$5.00$^c$ & 4200      \\
 Sun            & G2V    & --              & $-$4.908    & 4570      \\
 HD\,199288     & G2V    & 22.1$\pm$0.2    & $-$4.83$^e$ & 7700$^f$  \\
\hline
 HD\,59967      & G3V    & 21.8$\pm$0.2    & $-$4.43$^e$ & 100$^g$   \\
 HD\,33262      & F9V    & 11.65$\pm$0.03  & $-$4.65$^e$ & 250$^g$   \\
 HD\,165185     & G1V    & 17.6$\pm$0.2    & $-$4.54$^c$ & 291       \\
 HD\,20630      & G5V    & 9.14$\pm$0.02   & $-$4.420    & 522       \\
 HD\,97334      & G0V    & 21.9$\pm$0.2    & $-$4.422    & 551       \\
 HD\,106516     & F9V    & 22.4$\pm$0.4    & $-$4.651    & 1770      \\
 HD\,142373     & G0V    & 15.89$\pm$0.05  & $-$5.11$^c$ & 7400$^g$  \\
\hline
\end{tabular}
\end{center}
$^a$ -- From \citet{jarvinen2007}.\\
$^b$ -- From the WEBDA database \citep{webda}.\\
$^c$ -- From \citet{mamajek2008}.\\
$^d$ -- From \citet{melo2006}.\\
$^e$ -- From \citet{pace2013}.\\
$^f$ -- Average between \citet{holmberg2009} and \citet{casagrande2011}.\\
$^g$ -- From \citet{casagrande2011}.\\
$^h$ -- From \citet{figueira2014}.\\
$^i$ -- From \citet{murgas2013}.
\end{table}

\begin{figure*}
\begin{center}
\includegraphics[angle=-90,width=14.0cm,clip]{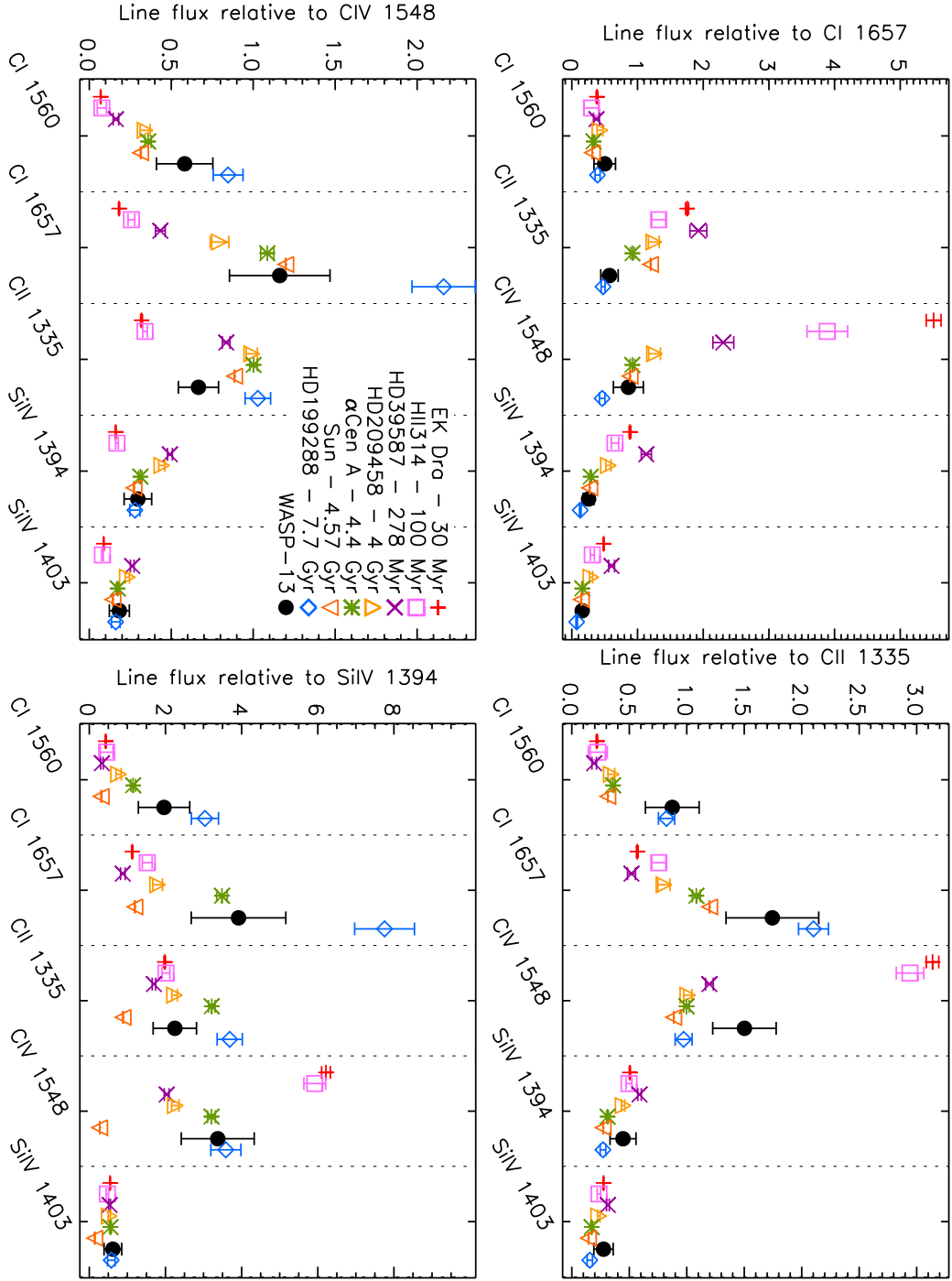}
\caption{Total flux of the measured features divided by the total flux of \ion{C}{1} $\lambda$1657 (top left), \ion{C}{2} $\lambda$1335 (top right), \ion{C}{4} $\lambda$1548 (bottom left), and \ion{Si}{4} $\lambda$1394 (bottom right) for WASP-13 (full black circle) and the comparison stars EK\,Dra (red plus sign), HII\,314 (pink square), HD\,39587 (purple cross), HD\,209458 (yellow upward-pointing triangle), $\alpha$\,Cen\,A (green asterisk), Sun (orange downward-pointing triangle), and HD\,199288 (blue diamond).}
\label{fig:lineratios}
\end{center}
\end{figure*}

The line flux ratios obtained for WASP-13 and the comparison stars are shown in Figure~\ref{fig:lineratios}. For the comparison stars, we derived the line fluxes using the same wavelength regions used for WASP-13 (Table~\ref{tab:wasp13-linefluxes}). For all stars, the ratio between the \ion{Si}{4} lines is about 0.5, in line with what is expected in the presence of an optically thin plasma \citep{math1999,bloomfield2002}.

For young and active stars, the features formed in the high-temperature transition region of the chromosphere (e.g., \ion{C}{4} and \ion{Si}{4}) are stronger than the low-temperature features \citep[e.g., \ion{C}{1} and \ion{C}{2};][]{wood1997}. The flux ratio of high- to low-temperature features decreases then with age (decreasing stellar activity). As expected, the young stars EK\,Dra and HII\,314 have the largest flux ratios of high- to low-temperature features (e.g., \ion{C}{4} $\lambda$1548/\ion{C}{1} $\lambda$1657 flux ratio in the top left panel of Figure~\ref{fig:lineratios}). Therefore, the comparisons shown in Figure~\ref{fig:lineratios} allow one to estimate the actual stellar activity and age of WASP-13. 

For the analysis of the results shown in Figure~\ref{fig:lineratios}, one has to consider that the \ion{C}{2} 1335 flux is strongly dependent on the amount of ISM absorption. For this reason, among the available comparisons, it is best to concentrate on those involving \ion{C}{1} $\lambda$1657, \ion{C}{4} $\lambda$1548, and \ion{Si}{4} $\lambda$1394. WASP-13 seems to have an FUV emission spectrum, and hence activity level, in between that of $\alpha$\,Cen\,A/Sun and HD\,199288, the oldest and most inactive comparison stars, with a slightly closer similarity to that of $\alpha$\,Cen\,A/Sun (see, e.g., \ion{C}{1} $\lambda$1657--\ion{C}{4} $\lambda$1548 flux ratio). This points toward an age of WASP-13 of about 5--6\,Gyr.

\begin{figure}
\begin{center}
\includegraphics[width=\hsize,clip]{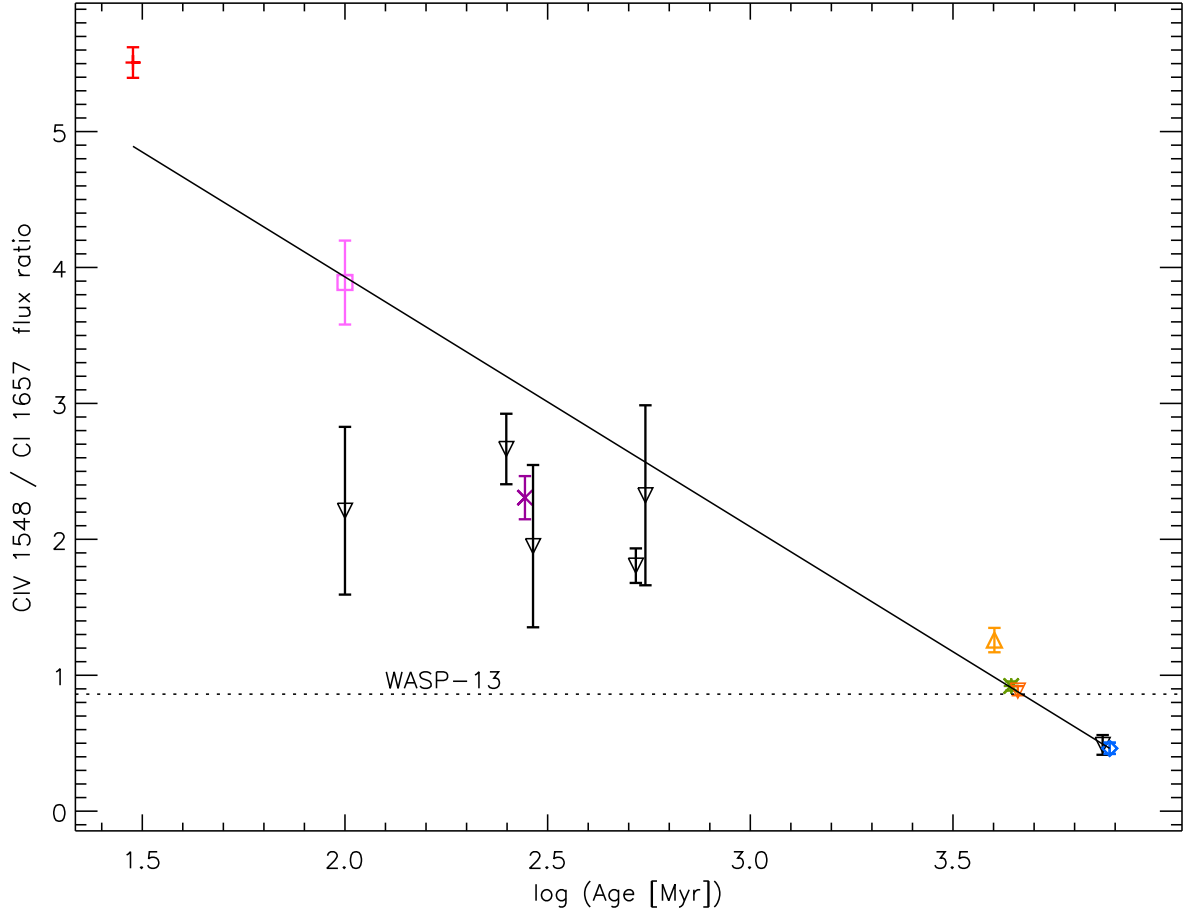}
\caption{\ion{C}{4} $\lambda$1548/\ion{C}{1} $\lambda$1657 flux ratio as a function of the logarithm of stellar age (in Myr) for the stars listed in Table~\ref{tab:comp.stars}. The black circles correspond to the stars listed in the second half of Table~\ref{tab:comp.stars}, while the other symbols are as in Figure~\ref{fig:lineratios}. The black solid line shows the linear fits in the log(age [Myr])--flux ratio plane. The black dashed line indicates the flux ratio derived for WASP-13.}
\label{fig:fit2lineratios}
\end{center}
\end{figure}

To best estimate the age of WASP-13, Figure~\ref{fig:fit2lineratios} shows the \ion{C}{4} 1548/\ion{C}{1} 1657 flux ratio as a function of stellar age. To increase the statistics and to make the age--flux ratio relation less dependent on the uncertainties on the ages and time-variable activity of the comparison stars, we added more comparison stars to the sample (see the second half of Table~\ref{tab:comp.stars}). To increase the number of stars, we eased the condition on the spectral type by including main-sequence stars between G5 and F9. The spectra, downloaded from the StarCAT catalog \citep{ayres2010}, were obtained with the E140M grating of the STIS. We measured the line fluxes as for WASP-13 and the other comparison stars.

We derived a linear relation in the log(age [Myr])--flux ratio plane, obtaining
\begin{equation}
\frac{\mathrm{CIV}\,1548}{\mathrm{CI}\,1657} = 7.607(\pm0.150) - 1.838(\pm0.041) \times \mathrm{log(Age\,[Myr])}\,\,,
\label{eq:age}
\end{equation}
with a Spearman-rank correlation coefficient of $-$0.913. From Equation~\ref{eq:age} and taking into account the uncertainties on the parameters of the fit and on the measured \ion{C}{4} $\lambda$1548/\ion{C}{1} $\lambda$1657 flux ratio, we derived an age of WASP-13 of 5.1$\pm$2.0\,Gyr, in agreement with what one can roughly estimate from Figure~\ref{fig:lineratios}.

The literature reports a rather wide age spread for WASP-13, ranging from 4 to 12.5\,Gyr \citep{skillen2009,barros2012,yilen2013,bonfanti2015}. All ages reported in the literature have been derived on the basis of different sets of stellar evolution tracks and atmospheric stellar parameters; hence, the large spread is not too surprising. The thorough spectral analysis shown by \citet{yilen2013} led to a clear improvement of the stellar atmospheric parameters over the previous set given by \citet{skillen2009}: their slightly hotter stellar effective temperature moved the star toward younger ages compared to the other estimations, which had all been based on the stellar atmospheric parameters given by \citet{skillen2009}. 

The age range derived here from the comparison of the FUV emission lines is closest to that given by \citet{yilen2013} of 4--5.5\,Gyr. This and the similarities of the FUV spectra of WASP-13 and $\alpha$\,Cen\,A/Sun, for which a rather precise age is available, increase our confidence in the age estimated from the chromospheric lines. By comparing the position of WASP-13 in the Hertzsprung--Russell diagram (HRD) to evolutionary tracks, \citet{yilen2013} showed that WASP-13 is very close to the turnoff point at the end of the main sequence, where stars stop burning hydrogen in the core. As a consequence of this, \citet{yilen2013} concluded that WASP-13 could equally be either a main-sequence star or a subgiant on its way to the Hertzsprung gap, but they did not consider the different speed of evolution in the two phases. The evolution of a star past the turnoff point, particularly in the contraction phase (small temperature increase after the turnoff point), is much quicker than any main sequence phase. This makes a main sequence evolutionary status for WASP-13 much more likely compared to a subgiant status; hence, WASP-13 is likely still burning hydrogen in the core.
\subsection{WASP-13's high energy flux}\label{sec:xuv}
The high-energy stellar irradiation, X-ray (1--100\,\AA) and EUV (100--912\,\AA), is one of the most critical parameters for exoplanet studies \citep[e.g.,][]{ribas2005,sanzforcada2011,linsky2014}, in particular for the analysis of the planet evaporation phenomenon and its evolutionary consequences \citep[e.g.,][]{davis2009,ehrenreich2011,mordasini2012,lopez2013}. For all stars beyond the solar system, directly measuring the complete XUV (X-ray\,$+$\,EUV) flux is not possible because the ISM is completely opaque in the EUV owing to absorption by neutral hydrogen. For this reason, various indirect methods and scaling relations have been derived to estimate the XUV flux of late-type stars on the basis of observables, such as X-ray fluxes \citep{sanzforcada2011,chadney2015}, Ly$\alpha$ fluxes \citep{linsky2014}, and stellar rotational velocities \citep{wood1994,colin2015,tu2015}.

Here we use a slightly different approach to derive the XUV flux of WASP-13. Given the similarity in the FUV line ratios between the Sun and WASP-13, we rescaled the solar irradiance reference spectrum \citep{woods2009} to match the observed WASP-13 flux of the \ion{C}{4} $\lambda$1548 feature, accounting for WASP-13's distance (see Section~\ref{sec:sed}) and radius \citep{yilen2013}. We used the \ion{C}{4} feature because it is expected to be unaffected by ISM absorption and hence the observed intensity can be directly compared to the Sun's. In addition, \ion{C}{4} has a formation temperature peaking near 10$^5$\,K, in the transition region, where much of the stellar EUV is generated \citep{linsky2014}. For this operation, we convolved the solar spectrum to the resolution of the COS spectrum and used the same integration limits (for both line and continuum) as those adopted for WASP-13. We then integrated the rescaled solar flux within the X-ray, EUV, and FUV wavelength ranges adopted by \citet[][their Table~4]{ribas2005} and \citet[][their Table~5]{linsky2014}, plus the Ly$\alpha$ flux (1212--1220\,\AA). The results are shown in Table~\ref{tab:EUVfluxes}. The method applied here should provide reliable results because WASP-13's photospheric temperature, metallicity, and FUV spectral characteristics are very similar to the Sun's, while the slightly larger mass should have a negligible effect compared to the uncertainties.
\begin{table}[ht]
\vspace{-2.0cm}
\caption[ ]{X-ray, EUV, and FUV fluxes at 1\,AU (in \ergscm) derived for WASP-13 by rescaling the solar spectrum (column two) and for the Sun itself (column three). The fluxes are calculated in the different bands adopted by \citet{ribas2005} and \citet{linsky2014}, plus various FUV bands including the Ly$\alpha$ line (1212--1220\,\AA). }
\label{tab:EUVfluxes}
\begin{center}
\begin{tabular}{l|ccc}
\hline
\hline
Wavelength  & \multicolumn{3}{|c}{Flux at 1\,AU} \\
Range (\AA) & \multicolumn{3}{|c}{(\ergscm)}     \\
            & \multicolumn{2}{|c}{WASP-13}        & Sun               \\
\hline
1--20                   & 0.031  &            & 0.013 \\
20--100                 & 0.374  &            & 0.150 \\
100--360                & 3.062  &            & 1.228 \\
360--920                & 1.978  &            & 0.794 \\
920--1180               & 1.379  &            & 0.553 \\
1--360$+$920--1180      & 6.622  &            & 2.657 \\
1--1180                 & 6.824  &            & 2.738 \\
\hline
100--200                & 1.124  &  1.599$^a$ & 0.451 \\
200--300                & 0.688  &  1.660$^a$ & 0.276 \\
300--400                & 1.365  &  1.902$^a$ & 0.548 \\
400--500                & 0.196  &  0.147$^a$ & 0.079 \\
500--600                & 0.332  &  0.534$^a$ & 0.133 \\
600--700                & 0.280  &  0.331$^a$ & 0.112 \\
700--800                & 0.286  &  0.739$^a$ & 0.115 \\
800--912                & 0.704  &  1.728$^a$ & 0.283 \\
912--1170               & 1.254  &  1.709$^a$ & 0.503 \\
\hline
1170--1300              & 15.893 & 	      & 6.377 \\
1300--1400              & 1.384  & 	      & 0.555 \\
1400--1500              & 1.391  & 	      & 0.558 \\
1500--1600              & 3.434  & 	      & 1.378 \\
1600--1700              & 8.066  & 	      & 3.236 \\
\hline
Ly$\alpha$ (1212--1220) & 14.497 & 15.860$^b$ & 5.817 \\
\hline
\end{tabular}
\end{center}
$^a$ -- From \citet{linsky2014} using the Ly$\alpha$ flux derived from the rescaled solar spectrum.\\
$^b$ -- From \citet{linsky2013}.\\
\end{table}


Table~\ref{tab:EUVfluxes} lists the integrated fluxes for WASP-13 and the Sun\footnote{Note that the solar fluxes are systematically slightly lower compared to those of \citet{ribas2005}, because the solar spectrum adopted here corresponds to that at minimum solar activity, while \citet{ribas2005} used a spectrum obtained at average solar activity.} at 1\,AU in various X-ray, EUV, and FUV wavelength regions. Despite the slightly lower activity of WASP-13 compared to that of the Sun, the solar fluxes at 1\,AU are smaller than those of WASP-13. This is due to the rather large difference in stellar radius between the two stars \citep{yilen2013}. As expected, a hypothetical WASP-13 with a radius of $R$\,=\,1\,\R\ would have XUV fluxes slightly smaller than the solar ones.

Table~\ref{tab:EUVfluxes} also lists the Ly$\alpha$ flux obtained using the relation given by \citet{linsky2013} and the EUV fluxes obtained using the relations given by \citet{linsky2014}. For almost all fluxes, the use of these relations leads to fluxes larger than those obtained from the scaling of the solar spectrum. The average difference is about 40\%, well within the expected uncertainties \citep[i.e., the EUV fluxes are based on a scaling relation of a scaling relation;][]{linsky2014}. This shows that, when possible, the use of a rescaled solar spectrum is to be preferred.

The Ly$\alpha$ line is clearly the most intense feature in the stellar high-energy spectrum with a total integrated flux of about 14\,\ergscm. The total XUV flux at 1\,AU, integrated within the 1--912\,\AA\ wavelength range, is about 5.3\,\ergscm, which at the orbital distance of WASP-13\,b is about 1850\,\ergscm. We used the scaled solar spectrum (see above) in the escape model of \citet{koskinen2013a,koskinen2013b} to estimate the mass-loss rate from WASP-13\,b. We placed the lower boundary of the model at a pressure $p$\,=\,10$^{-6}$\,bar, where the radius is $r$\,=\,1.15\,$R_{\rm p}$, based on the temperature of 1500\,K \citep{barros2012} in the lower atmosphere. In the absence of any information on the presence of heavier species in the escaping atmosphere, the model includes only H$_2$, H, and He with their associated ions H$^+$, H$_2^+$, H$_3^+$, He$^+$, and HeH$^+$. Because H$_2$ dissociates just above the lower boundary, the model is practically composed of H, H$^+$, and electrons above 1.2\,$R_{\rm p}$.  

The model predicts a mass-loss rate of 1.5\,$\times$\,10$^{11}$\,g\,s$^{-1}$ (2.5\,$\times$\,10$^{-12}$\,$M_{\rm J}$\,yr$^{-1}$). We compared this with the energy-limited mass-loss rate, written as \citep[e.g.,][]{erkaev2007}
\begin{equation}
\dot{M} = \frac{\eta_E \pi r_E^2 F_E}{K_s \Phi_0}
\end{equation}  
where $\eta_E$ is the mass loss efficiency, $r_E$ is the effective radius where stellar energy is absorbed, $1/K_s$ is the enhancement factor due to Roche lobe effects, $F_E$ is the wavelength-integrated stellar flux, and $\Phi_0 = G M_p / R_0$ is the surface gravitational potential. The mass-loss rate predicted by the energy-limited formula (we adopted $\eta_E$\,=\,0.46, $r_E$\,=\,1.75\,$R_p$, and $1/K_s$\,=\,1.15) is within 1\% of that given by the hydrodynamic model. We note that the mass loss rate we obtained from the hydrodynamic model is about a factor of 10 higher than the mass-loss rate of 2.5\,$\times$\,10$^{10}$\,g\,s$^{-1}$ predicted by \citet{ehrenreich2011} on the basis of the energy-limited formulation. This is because the surface gravity of WASP-13\,b is relatively low, leading to a rather high altitude for the XUV heating peak, and because \citet{ehrenreich2011} assumed a lower mass-loss efficiency of $\eta_E$\,=\,0.15. Our value of $\eta_E$\,=\,0.46 is based on a photoelectron heating efficiency of 0.93 that should be suitable for EUV photons under strong ionization \citep{cecchi09,koskinen2013a,koskinen2013b} and cooling by recombination. It is also reasonably close to the thermospheric heating efficiency on many solar system atmospheres.

The mass-loss rate is not large enough to substantially decrease the planet's mass within the main-sequence lifetime of the star. Nevertheless, the mass-loss rate is as high as that derived by \citet{ehrenreich2011} for WASP-12\,b, which is about 1000\,K hotter than WASP-13\,b and for which translucent circumstellar material appears to be present\footnote{The previous mass-loss rate derived for WASP-12 on the basis of the energy-limited formula may be underestimated (Koskinen et al. 2016, in preparation).}. One may therefore expect the same for the WASP-13 system, although additional factors, such as the stellar magnetic field configuration \citep{lanza2014} and stellar wind/radiation, may also play a role.
\subsection{Circumstellar material}\label{sec:cloud}
The analysis of the COS spectrum has shown that WASP-13 has a chromosphere typical of Sun-like stars, in contrast to what was indicated by the very low \logR\ value, if taken as reflecting the intrinsic stellar properties. For WASP-13 we can therefore definitely exclude the possibility that the planet has significantly affected stellar activity, as \citet{pillitteri2014} suggested for WASP-18. The hot Jupiter WASP-18\,b is 20 times more massive and has a semi-major axis 2.5 times smaller than WASP-13\,b, so this plausibly explains different situations in the two systems. If WASP-13\,b was capable of reducing the activity of the host star, many more systems would be expected to present a very low \logR\ value. This in turn implies that extrinsic absorption, either from the ISM and/or from material local to the system, is likely to be the cause of the \logR--FUV flux discrepancy for the majority of planet hosts sharing it.

\begin{figure}
\begin{center}
\includegraphics[width=\hsize,clip]{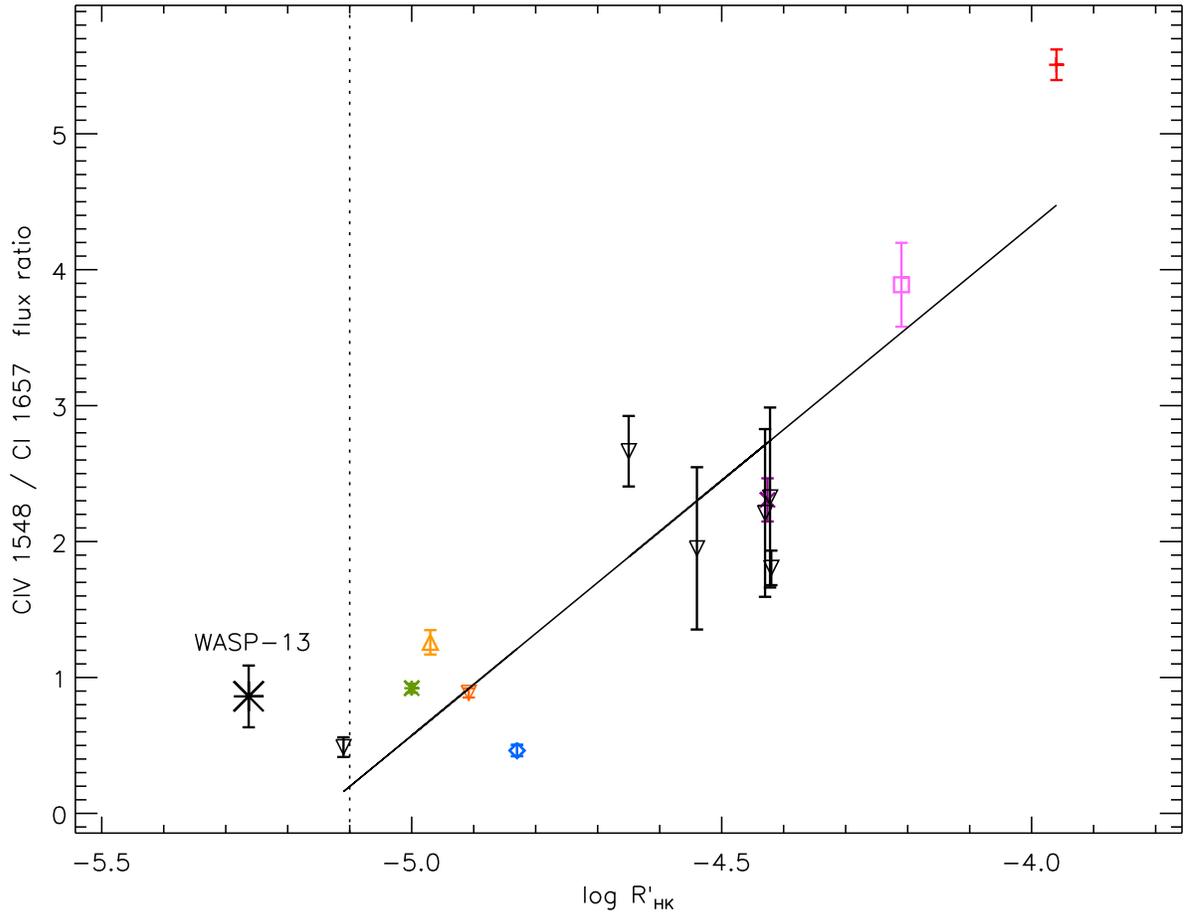}
\caption{\ion{C}{4} $\lambda$1548/\ion{C}{1} $\lambda$1657 flux ratio as a function of the \logR\ for the stars listed in Table~\ref{tab:comp.stars} and WASP-13. The black circles correspond to the stars listed in the second half of Table~\ref{tab:comp.stars}, while the other symbols are as in Figure~\ref{fig:lineratios}. WASP-13 is marked by a large black asterisk. The black solid line shows the linear fit in the \logR--flux ratio plane obtained without considering WASP-13. The black dashed line indicates the \logR\ basal level \citep{wright2004}.}
\label{fig:fit2lineratioslogR}
\end{center}
\end{figure}

The presence of absorption biasing the measurement of the \logR\ value appears clear when relating the \ion{C}{4} $\lambda$1548/\ion{C}{1} $\lambda$1657 flux ratio and the \logR\ value for WASP-13 and the comparison stars (Figure~\ref{fig:fit2lineratioslogR}). From a linear fit performed without considering WASP-13, we obtained the following correlation:
\begin{equation}
\frac{\mathrm{CIV}\,1548}{\mathrm{CI}\,1657} = 19.332(\pm2.508) + 3.752(\pm0.543) \times \mathrm{\log R^{\prime}_{\mathrm{HK}}}\,\,.
\label{eq:logR}
\end{equation}

Figure~\ref{fig:fit2lineratioslogR} and Equation~\ref{eq:logR} indicate that the true \logR\ value of WASP-13 should be about $-$4.95, similar to the solar value and well above the measured one of $-$5.263. This highlights even further the presence of a bias in the measured \logR\ value, which is probably caused by extrinsic absorption by the ISM and possibly by circumstellar material.

\begin{figure}
\begin{center}
\includegraphics[width=\hsize,clip]{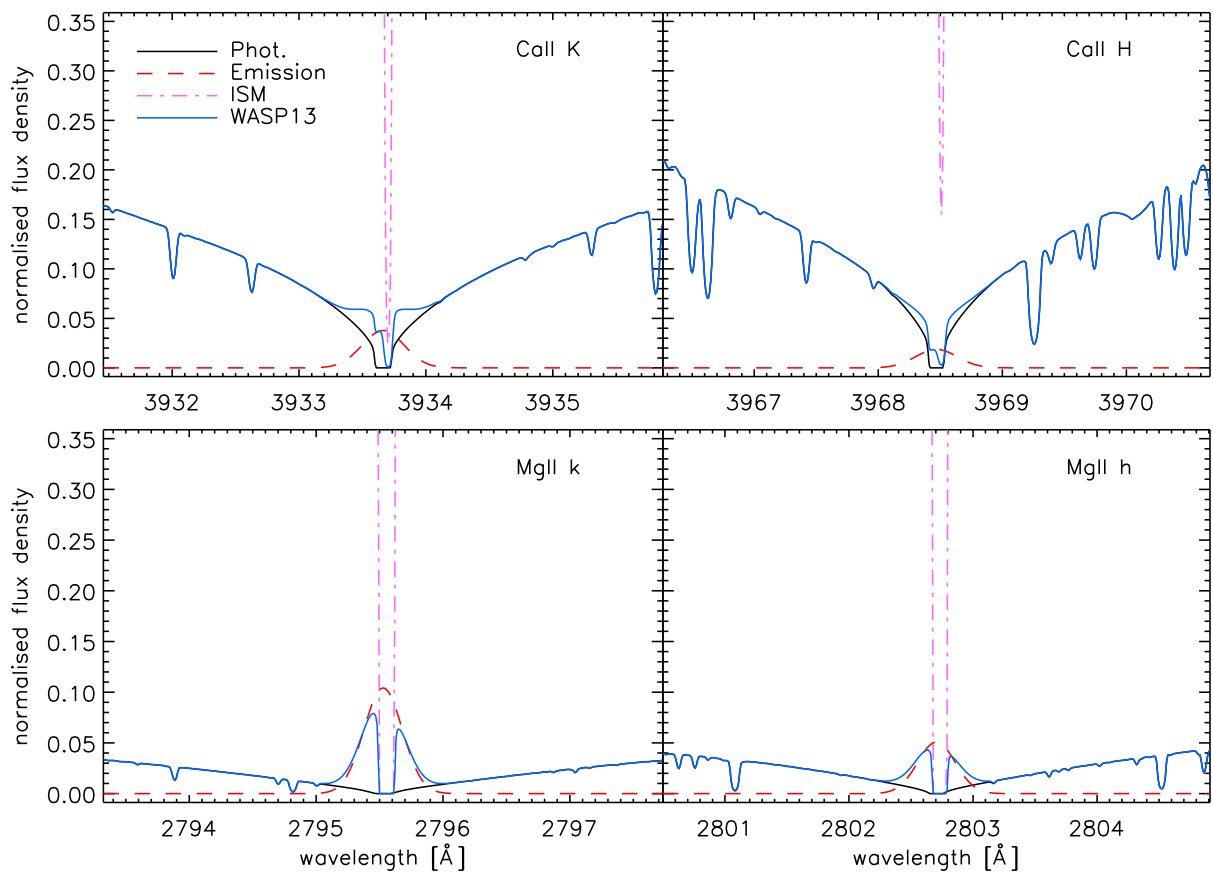}
\caption{Step-by-step construction of the synthetic profiles at the core of the \ion{Mg}{2}\,h\&k (bottom) and \ion{Ca}{2}\,H\&K (top) resonance lines. The black solid line shows the \llm\ photospheric spectrum, the red dashed line shows the line core emission, while the purple dot-dashed line shows the ISM absorption, shifted by $+$3.0\,\kms\ to match the position of the ISM absorption component observed in the spectrum of WASP-13 (see Figure~\ref{fig:em-abs-comp}). The blue solid line shows the final composite synthetic spectrum. Profiles do not include instrumental broadening.}
\label{fig:em-abs}
\end{center}
\end{figure}

To look for the possible presence of a spectral signature of absorption local to the system, hence additional to that of the ISM, we compared the observed HIRES spectrum\footnote{The reduced HIRES spectrum of WASP-13, having a resolving power of R\,$\sim$\,48,000, was downloaded from the Keck archive {\tt https://koa.ipac.caltech.edu/cgi-bin/KOA/nph-KOAlogin}.} of WASP-13 with a synthetic spectrum of the star, which includes the chromospheric emission and ISM absorption components, in the region of the \ion{Ca}{2}\,H\&K lines. We estimated the total \ion{Ca}{2} line core emission using the Ly$\alpha$--\ion{Ca}{2}\,H\&K relations given by \citet{linsky2013} and distributed the total emission flux between the two lines according to their oscillator strengths (i.e., \loggf\ values). For this operation we considered the Ly$\alpha$ flux we obtained from the rescaling of the solar spectrum namely, 14\,\ergscm\ at 1\,AU, obtaining a total \ion{Ca}{2} emission flux of 19\,\ergscm\ at 1\,AU. We simulated the chromospheric line emission using Gaussian profiles broadened in order to match the line width of the solar emission line profiles. The ISM absorption was modeled using Voigt profiles with a \ion{Ca}{2} column density derived using the ionization mix by \citet{frisch2003} and assuming a broadening $b$-parameter of 2.0\,\kms\ \citep{redfield2002}. Although an NUV spectrum of WASP-13 is not available, following the same procedure, we also calculated synthetic spectra in the region of the \ion{Mg}{2}\,h\&k lines \citep[following][we obtained a total \ion{Mg}{2} emission flux of 30\,\ergscm\ at 1\,AU]{linsky2013}. The photospheric synthetic spectrum of WASP-13 was calculated with \llm\ \citep{shulyak2004}. Figure~\ref{fig:em-abs} shows the synthetic profiles derived assuming a logarithmic \ion{Mg}{2} column density of 15.00 and a logarithmic \ion{Ca}{2} column density of 12.00; see Table~\ref{tab:ebv-columnDensities}.

\begin{figure}
\begin{center}
\includegraphics[width=\hsize,clip]{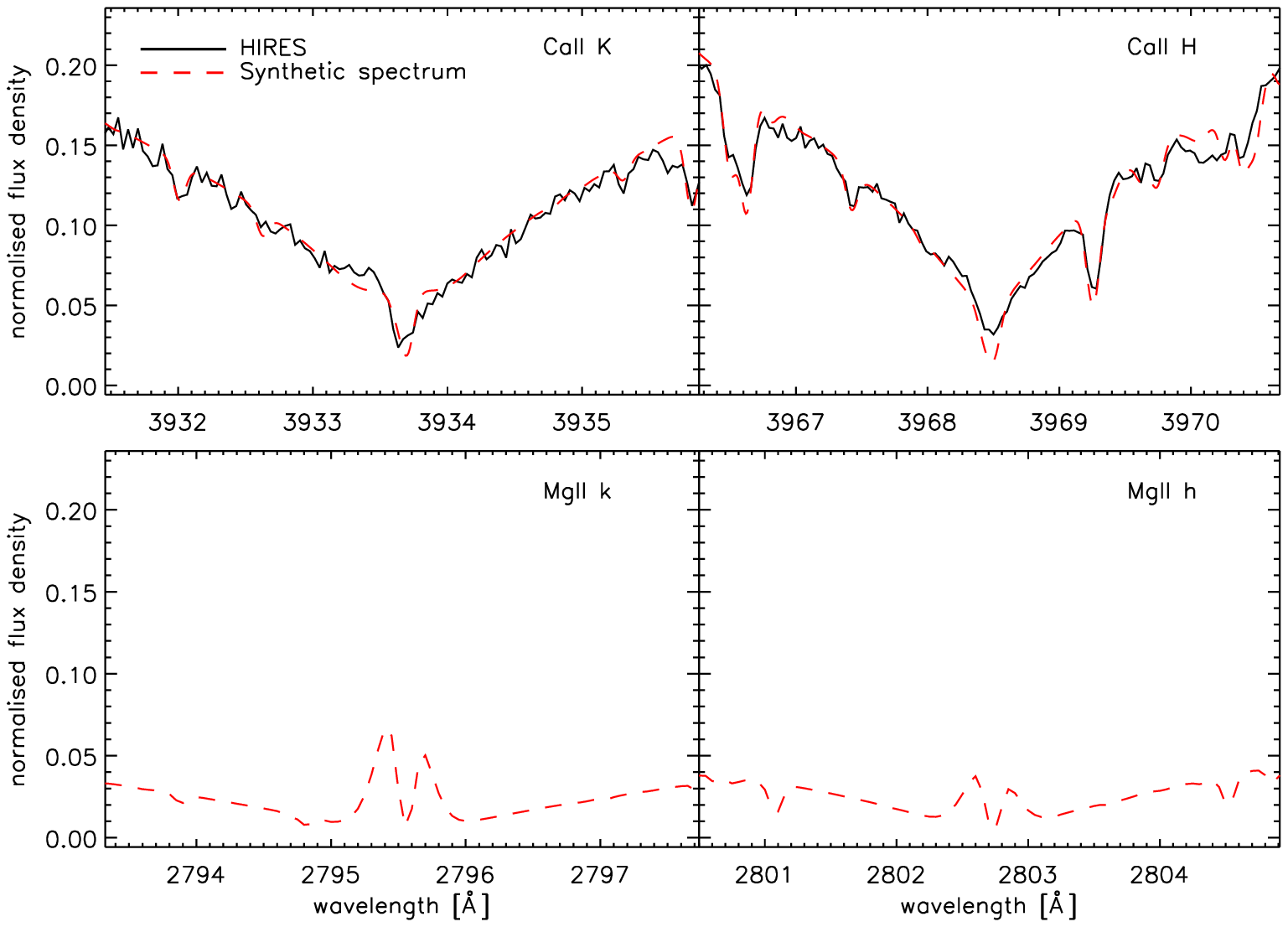}
\caption{Top: comparison between the HIRES spectrum (black solid line) of WASP-13 in the region of the \ion{Ca}{2}\,H\&K line cores and a synthetic spectrum (red dashed line), which includes chromospheric emission and ISM absorption adopting the parameters given in the text. Bottom: synthetic spectrum of WASP-13, including chromospheric emission and ISM absorption, convolved with the resolution and sampling of the STIS E230M grating.}
\label{fig:em-abs-comp}
\end{center}
\end{figure}

Figure~\ref{fig:em-abs-comp} shows the comparison of the synthetic spectrum, obtained including chromospheric emission and ISM absorption with the observed HIRES spectrum, and the expected \ion{Mg}{2}\,h\&k line core profiles of WASP-13 convolved with the resolution and sampling of the STIS E230M grating. As expected, the lower ISM column density, derived using the ionization mix by \citet{frisch2003}, provides a better fit to the observed spectrum, compared to that obtained using solar abundances. The synthesized emission is slightly lower than the observed emission, but a simple increase of the emitted \ion{Ca}{2} flux would not lead to a significantly better fit because it would exacerbate the mismatch on the blue side of the ISM absorption (see \ion{Ca}{2}\,K line).

On the basis of the assumed characteristics of the ISM along the WASP-13 line of sight, Figure~\ref{fig:em-abs-comp} would indicate that absorption from material local to the system is not present. Nevertheless, the ISM \ion{Ca}{2} column density is too poorly constrained to firmly exclude the presence of extra absorption (i.e., additional to the ISM absorption) in the line core. Note also that we assume the presence of a single ISM absorption component; this may not be the case. The bottom panel of Figure~\ref{fig:em-abs-comp} shows that assuming that the only nonphotospheric absorption present in the spectrum is due to the ISM, WASP-13 would indeed present \ion{Mg}{2} line core emission that would be detectable with STIS, contrary to the case of WASP-12 where no emission was detected \citep{haswell2012}. An NUV spectrum of WASP-13 is needed to look for the presence of \ion{Mg}{2} line core emission and compare it with our prediction.

Much better knowledge of the ISM absorption component(s) is needed to disentangle the ISM absorption from the circumstellar absorption, if present. This may be obtained from the measurement of the ISM absorption from early-type stars lying close to the WASP-13 line of sight, following the work of \citet{fossati2013} for the WASP-12 system. This would allow us to accurately measure ISM absorption parameters and directly derive an unbiased value of \logR. Thus, the idea that the ISM alone could cause the low \logR\ can be tested. The major problem in this procedure would be that the distance to the early-type stars would not be known and would almost certainly not match that of WASP-13. WASP-13 lies far from the Galactic plane, in a sparsely populated region of the sky with no nearby early-type stars. To offset this disadvantage, the ISM absorption is mostly confined to the Galactic plane, so closely matched distances are less critical for high Galactic latitude lines of sight. Further attempts to identify the spectroscopic signature of the extra absorption could be made on the basis of solid theoretical estimates of the temperature of the absorbing material, though the few values currently available \citep[10$^3$--10$^5$\,K;][]{li2010,lanza2014} are too poorly constrained to serve this purpose.
\section{Conclusions}\label{sec:conclusions}
Like other stars hosting hot Jupiters, exemplified by WASP-12, the G1V planet-hosting star WASP-13 presents a \logR\ value well below the basal level of solar-like main-sequence stars. The presence of absorption in the core of the resonance lines often used to measure stellar activity has been attributed to ({\it i}) the ISM, ({\it ii}) material local to the system, or ({\it iii}) a combination of local and ISM absorption. A direct influence of the planet on the stellar dynamo mechanism has also been suggested.

In 2015 April, we obtained HST/COS FUV spectra of WASP-13, timed with the planet's primary transit. We have measured the actual stellar activity from the high-ionization lines free from extrinsic absorption, estimated the high-energy flux, and assessed the absorption from material local to the system. Measurements of the high-energy flux for these very inactive stars are particularly important: on the basis of the activity parameter alone, it may not be possible to reliably infer the XUV stellar flux. This XUV flux irradiates the planet and is needed as input by models of the planetary atmospheres' response to the intense irradiation from the host star. We have demonstrated that a modest investment of HST time allows us to retrieve the stellar XUV flux and derive, e.g., the planetary mass-loss rate. Without this, evolution calculations for these hot Jupiters are speculative and subject to large uncertainties.

We integrated the stellar flux in the FUV wavelength band dominated by the stellar continuum, but the uncertainties were too large to detect the planet transit. We calculated the expected transit depth in the considered wavelength region assuming Rayleigh scattering, concluding that it should be 9\% deeper than the optical transit. On the basis of the COS observations, we did not detect at 2$\sigma$ Rayleigh scattering in the planet atmosphere.

We analyzed the star's SED spanning from the UV to the IR. On the basis of the stellar effective temperature and radius, we estimated a distance to the star of $d$\,=\,232$\pm$8\,pc and a reddening of \ebv\,=\,0.045$\pm$0.25\,mag. Using various relations of the \ion{H}{1} column density with \ebv, we estimated the ISM logarithmic column density of \ion{H}{1}, \ion{Mg}{2}, \ion{Ca}{2}, and \ion{Na}{1} obtaining on average 20.50, 15.00, 12.00, and 12.20, respectively.

From the co-added COS spectrum we measured the integrated flux for the \ion{C}{1} multiplets at $\sim$1560\,\AA\ and $\sim$1657\,\AA, the \ion{C}{2} doublet at $\sim$1335\,\AA, the \ion{C}{4} doublet at $\sim$1548\,\AA, and the \ion{Si}{4} doublet at $\sim$1394\,\AA\ and $\sim$1403\,\AA. No further stellar lines appeared to be strong enough to be visible in the spectrum. We detected at $\gtrsim$4$\sigma$ lines belonging to three different ionization states of carbon (the \ion{C}{1} multiplet at $\sim$1657\,\AA, the \ion{C}{2} doublet at $\sim$1335\,\AA, and the \ion{C}{4} doublet at $\sim$1548\,\AA) and the \ion{Si}{4} line at $\sim$1560\,\AA\ at $\sim$3$\sigma$. These detections clearly indicate that, despite the low stellar activity index, WASP-13 hosts a chromosphere consistent with that of other cool stars (e.g., the Sun) and that the star is not completely inactive. The stellar dynamo mechanism is therefore not significantly affected by the planet.

We compared the line flux ratios derived for WASP-13 with those of main-sequence early G-type stars of different ages and activity levels. We see a clear decrease of the flux ratios of high- to low-temperature features (e.g., \ion{C}{4} 1548/\ion{C}{1} 1657 flux ratio) with age. By considering the flux ratios calculated from the lines less affected by ISM absorption (\ion{C}{1} $\lambda$1657, \ion{C}{4} $\lambda$1548, and \ion{Si}{4} $\lambda$1394), WASP-13 appears to be only slightly older and more inactive than the Sun and $\alpha$\,Cen\,A. We concentrate on the \ion{C}{4} $\lambda$1548/\ion{C}{1} $\lambda$1657 flux ratio and derive a linear log(age [Myr])--flux ratio relation, from which we estimate an age of WASP-13 of 5.1$\pm$2.0\,Gyr.

Given the similarities between the FUV spectra of WASP-13 and the Sun, we rescaled the solar irradiance reference spectrum to match the flux of the \ion{C}{4} $\lambda$1548 doublet, thus estimating the stellar high-energy flux, taking into account the distance and radius of WASP-13. By integrating the rescaled solar spectrum in the 1--912\,\AA\ wavelength range, we obtained a flux of 5.4\,\ergscm\ at a distance of 1\,AU and 1859\,\ergscm\ at the distance of WASP-13\,b. We used the rescaled solar spectrum as input for a detailed hydrodynamic modeling of the planet's upper atmosphere, deriving a mass-loss rate of 1.5$\times$10$^{11}$\,g\,s$^{-1}$ (2.5\,$\times$\,10$^{-12}$\,$M_{\rm J}$\,yr$^{-1}$). This value is about 10 times higher than previous estimates based on the energy-limited formula.

For WASP-13, the \logR\ value of $-$5.263 is clearly not a good measure of the stellar activity, which we find to be only slightly lower than that of the Sun, which has an average \logR\ value of $-$4.908. This strongly suggests the presence of absorption in the core of the \ion{Ca}{2}\,H\&K resonance lines used to measure the \logR\ value. 

We looked for the absorption signature of veiling gas in a high-resolution Keck/HIRES spectrum of WASP-13 covering the \ion{Ca}{2}\,H\&K line cores. We derived the expected \ion{Ca}{2}\,H\&K line core emission and modeled it together with the photospheric spectrum and the ISM absorption. The composite spectrum was then compared to the observation. The large uncertainty in the \ion{Ca}{2} ISM abundance did not allow us to arrive at any firm conclusion on the presence of absorption additional to that of the ISM. In the case of WASP-13, the ISM may provide the only cause of the biased \logR\ value. A high-resolution spectrum covering the core of the \ion{Mg}{2}\,h\&k resonance lines is needed to establish a firm solution for the origin of the low \logR\ value of WASP-13.
\section*{Acknowledgments}
This work is based on observations made with the NASA/ESA Hubble
Space Telescope, obtained from MAST at the Space Telescope Science Institute,
which is operated by the Association of Universities for Research in
Astronomy, Inc., under NASA contract NAS 5-26555. These observations are
associated with program no. 13859, to which support was provided by NASA through
a grant from the Space Telescope Science Institute. This publication makes use
of data products from the Two Micron All Sky Survey, which is a joint project
of the University of Massachusetts and the Infrared Processing and Analysis
Center/California Institute of Technology, funded by the National Aeronautics
and Space Administration and the National Science Foundation. Support for StarCAT was provided by grant HST-AR-10638.01-A from STScI and grant NAG5-13058 from NASA. This work has made use of public databases hosted by SIMBAD and VizieR, both maintained by CDS, Strasbourg, France. L.F. acknowledges financial support from the Alexander von Humboldt foundation. C.H. is supported by STFC under grant ST/L000776/1. We thank the anonymous referee for the useful comments.
{\it Facilities:} \facility{HST (COS)},  \facility{HST (STIS)}, \facility{Keck (HIRES)}.

\end{document}